\documentclass{aa}  

\usepackage{graphicx}
\usepackage{natbib}
\usepackage{color}
\usepackage{upgreek}
\usepackage{soul}

\newcommand{\bm}{\langle B\rangle}
\newcommand{\bz}{\langle B_{\rm z}\rangle}

\bibpunct{(}{)}{;}{a}{}{,}

\usepackage{txfonts}
\usepackage{mathrsfs}
\begin{document}

\titlerunning{HD\,57372}
\authorrunning{Hubrig et al.}

\title{Extreme Magnetic Field Modulus Variability of the Bp star HD\,57372}

   \author{
	  S.~Hubrig\inst{1}
          \and
          S.~D.~Chojnowski\inst{2}
          \and
          S.~P.~J\"arvinen \inst{1}
          \and
	  I.~Ilyin\inst{1}
          \and
          K.~Pan\inst{3}
          }

   \institute{Leibniz-Institut f\"ur Astrophysik Potsdam (AIP),
              An der Sternwarte~16, 14482~Potsdam, Germany\\
              \email{shubrig@aip.de}
           \and NASA Ames Research Center, Moffett Field, CA 94035, USA
           \and Apache Point Observatory and New Mexico State
              University, P.O. Box 59, Sunspot, NM, 88349-0059, USA}
    \date{Received MM DD, 2024; accepted MM DD, 202X}
 \abstract
      {In chemically peculiar Ap/Bp stars with large-scale organised magnetic fields with a simple centred dipole configuration,
the ratio between the maximum and the minimum
of the mean magnetic field modulus is of the order of 1.25.
Values of 2 or more are observed only for very few Ap/Bp stars and are indicative of a very unusual
magnetic field geometry.
 }
{Determining the magnetic field structure of Ap/Bp stars
is bound to provide a different insight into the physics
and the origin of the magnetic fields in early-type stars.
 In this respect, the Bp star HD\,57372 is of particular interest because
strongly variable magnetically split lines are observed in HARPS and APOGEE spectra. 
}
   {We obtained and analysed measurements of the mean magnetic
field modulus and of the mean longitudinal magnetic field using near-infrared spectra and
optical polarimetric spectra distributed over the stellar rotation period.
}
   {The mean magnetic field modulus $\bm$ of HD\,57372, as estimated from absorption lines that are split via the
     Zeeman effect and resolved in both optical and near-infrared spectra, is found to vary by an extraordinary
     amount of about 10\,kG.
The exceptional value of 3 for the ratio between the maximum and the minimum
of the field modulus is indicative of a very unusual geometry of HD\,57372's magnetic field.
All observable quantities are found to vary in phase with the photometric period of 7.889\,day.
This includes the longitudinal magnetic field $\bz$, which varies from $-6$\,kG up to $1.7$\,kG in FORS2 spectra
as well as the metal line strengths, whose equivalent widths change by up to 50\% of their mean values
over the course of the rotation period.
The B8 temperature class of HD\,57372 also places it among the hottest stars known to exhibit resolved,
magnetically split lines.
}
{}

   \keywords{
stars: magnetic fields --
stars: chemically peculiar --
stars: early-type --
stars: variables: general
               }

   \maketitle
%

\section{Introduction}
\label{sect:intro}

For well over 50 years it has been known that a significant fraction of late-B and A type stars host strong,
large-scale magnetic fields that lead to peculiar surface abundances and hence also to their classification
as chemically peculiar (CP) Ap/Bp stars \citep{babcock1958}.
These stars are slow rotators as a rule when compared to their non-magnetic analogues,
and in the most extreme cases of slow rotation, strong magnetic fields,
and the presence of magnetically sensitive spectral lines,
it is possible to resolve the individual magnetically split components in unpolarized light.
Measurement of the wavelength separations of the split components combined with knowledge
of the associated quantum numbers gives a very simple yet extremely valuable measure
of the intrinsic surface magnetic field strength as averaged over the visible hemisphere,
a quantity known as the magnetic field modulus \citep[$\bm$; see][]{mathys1990,mathys1992,mathys1997,mathys2017}.
For all other magnetic stars, diagnosis of the intrinsic magnetic field strength usually relies
on spectropolarimetric observations of circular polrization over the full rotation period.

The first example of resolved, magnetically split lines (RMSL) in an Ap/Bp star
was HD\,215441 \citep[more commonly known as Babcock's Star;][]{babcock1960},
and despite more than a hundred additional examples having subsequently been found,
Babcock's Star remains unique to this day.
Not only is it the most magnetic main sequence star known ($\bm\sim$34\,kG),
it is a relatively rapid rotator (compared to most RMSL stars) and also the hottest star
for which RMSL have been observed.
In particular, \citet{preston1969} used time series $\bm$ measurement to confirm
that the 9.49\,days photometric period found by \citet{Jarzebowski1960}
was in fact the rotational period ($P_{\rm rot}$) of Babcock's Star,
and based on the strength of temperature sensitive features in the spectra,
he concluded that the temperature class could be as hot as B3 or B4.

In contrast to Babcock's Star, the majority of Ap/Bp stars known to exhibit RMSL have far longer rotation periods.
For example, the 29\,year rotation of HD\,50169 could only be determined
after a few decades of observation \citep{mathys2019},
and the rotation period of HD\,101065 \citep[also known as Przybylski's Star;][]{przybylski1961}
could only be constrained to $\sim188$\,years by \citet{Hubrig2018} despite an even longer observational baseline.
With early/mid-A spectral types, typical RMSL stars are also cooler than Babcock's Star,
and with $\bm$ usually less than 10\,kG, they are also considerably less magnetic \citep{mathys2017}.

The most comprehensive studies of RMSL stars to date have been provided by \citet{mathys2017}
in particular essentially providing an encyclopedia on the magnetic properties of what at the time was an 84-star sample.
Among the key findings of this work was confirmation of an already-suspected dearth of short-period binaries
among the RMSL stars, leading to the suspicion that the peculiarities of these stars may stem from binary mergers.
In lieu of a theory for generating a magnetic fields in a purely radiative atmosphere,
the binary merger hypothesis is a promising one that has only recently gained traction
to the point where a theoretical basis is now being developed \citep[e.g.][]{Schneider2020}.

For numerous RMSL stars, both the $\bm$ and the brightnesses of the stars
vary as a function of rotation phase due to the changing viewing angle of the magnetic poles,
such that rotation periods can be determined from either photometry or spectroscopy.
These variations can often be successfully reproduced under the assumption of a simple dipole
using the framework developed by \citet{preston1969} in his study of Babcock's Star,
whereby the ratio of $q=\bm_{\rm max}/\bm_{\rm min}$ should be of the order of 1.25 in the case of a pure dipole.
For some of the RMSL stars however, the magnetic field geometry is clearly more complicated.
Out of a subset of 28 RMSL stars with known rotation periods over which $\bm$ had been measured,
\citet{mathys2017} found eight stars with $q>1.25$, six of which had $q>1.3$.
Among these six stars, the A3p star HD\,65339 stood out as exceptionally remarkable
considering $P_{\rm rot}=8.03$\,days, $q\sim2.27$, and $\Delta_{\langle B\rangle}=\bm_{\rm max}-\bm_{\rm min}\sim9.7$\,kG \citep[see also][]{mathys1997}.
In terms of $\Delta_{\langle B\rangle}$, the only fair comparison currently known
is the A2p star HD\,126515, for which $\Delta_{\langle B\rangle}\sim6$\,kG,
albeit it over a much longer timescale of $P_{\rm rot}\sim130$\,days.

In the context of $q$, it is worth mentioning HR\,465, which has long been known
as one of the most extreme Bp stars in respect to its exotic atomic line content \citep[e.g.][]{cowley1975},
complicated variability over its long $\sim21$\,year rotation period \citep{Scholz1978},
and an unprecedentedly large $q\sim3.8$ \citep{mathys1997,giarrusso2022}.
As with Babcock's Star, HR\,465 is thus a severe outlier among the RMSL stars,
though in this case representing the extreme low end of all measured magnetic field
moduli, with $\bm$ $<3$\,kG on average.

Here, we present a new example of a RMSL Bp star whose magnetic properties are every bit
as exceptional as those of Ap stars such as HD\,65339 and HR\,465.
Prior to this work, HD\,57372 had not been specifically mentioned in any papers
beyond inclusion in the \citet{Bernhard2015} search for photometric variability in CP stars.
These authors reported a $V$-band magnitude that varies between 7.90--7.93 over the course of a 7.888\,day period.
The B8p\,Si spectral type, indicating enhanced Si features in the optical, was reported in the catalog of CP stars
by \citet{RensonManfroid2009}.

Four medium-resolution $H$-band spectra of HD\,57372 have been obtained by the northern Apache Point Observatory
Galactic Evolution Experiment \citep[APOGEE;][]{Majewski2017,Wilson2019} spectrograph, which operates as a sub-component of fifth instalment of the Sloan Digital Sky Survey \cite[SDSS-V; e.g.][and Kollmeier, J.A., in preparation]{Ahumada2020}.
Intriguingly, our inspection of these spectra shows extreme short-term variability of RMSL. To investigate HD\,57372 in more detail, we acquired eight high-resolution polarimetric spectra
from the High Accuracy Radial Velocity Planet Searcher \citep[HARPS;][]{Pepe2000,Mayor2003}
instrument attached to the European Southern Observatory (ESO) 3.6\,m telescope and twelve low-resolution
polarimetric spectra
from the FOcal Reducer/low dispersion Spectrograph 2 (FORS2) mounted on the 8.2-m Antu unit of the Very Large Telescope (VLT). Further, numerous observations of HD\,57372 by the Transiting Exoplanet Survey Satellite
\citep[TESS;][]{Ricker2015} are also available, and we make use of them for refining the rotation period.

The paper is laid out as follows.
Section~\ref{sect:spectro} describes the spectroscopic and photometric
data including the TESS lightcurve and associated refinement of $P_{\rm rot}$.
In Section~\ref{sect:gaia} we discuss the Gaia DR3 parameters and
in Section~\ref{sect:mfield} the magnetic field variability of HD\,57372 in both the optical
and the near-infrared (NIR)  $H$-band is presented. In Section ~\ref{sect:stellar_pars} we
explore stellar parameters assessed based on the Balmer line H$\delta$ and surface abundances of
HD\,57372 in comparison to other magnetic Ap/Bp stars.
Finally, in Section~\ref{sect:discussion}, we compare HD\,57372
with Ap/Bp stars known to exhibit magnetically resolved lines and discuss its magnetic field structure.

\section{Spectroscopy and photometry}
\label{sect:spectro}

\subsection{SDSS 2.5m$/$APOGEE}
\label{sect:apogee}
HD\,57372 was observed by the APOGEE spectrograph on the Sloan 2.5-m telescope \citep{Gunn2006}
at Apache Point Observatory (APO) four times over the course of a week between 2020 December~29 and 2021 January~6.
The spectra have a resolution of about 22\,500 and cover most of the $H$-band (15145--16960\,\AA{}; vacuum wavelengths
used throughout this paper when referring to the $H$-band) onto three detectors,
with gaps between 15800--15860\,\AA{} and 16430--16480\,\AA{} due to non-overlapping wavelength coverage of the detectors.
Total exposure times ranged from 49 to 82 minutes, and the associated signal-to-noise ratios ($S/N$) range from 230 to 264.
The reduction was carried out using the standard APOGEE data reduction pipeline described in \citet{Nidever2015},
with the only post-reduction steps being continuum normalization via fitting low-order polynomials to line-free regions,
correction of the wavelengths to rest frame based on Gaussian fitting of narrow absorption lines,
and manual removal of otherwise distracting airglow emission residuals. When analysing
individual absorption lines, we also re-normalised the local continuum
in small regions enclosing the lines.

\subsection{ESO 3.6m$/$HARPSpol}
\label{sect:harps}

HD\,57372 was observed with HARPS in the spectropolarimetric mode eight times between 2022 April~23 and 2022 May~1.
In the spectropolarimetric mode of the
HARPS has a resolving power of about 115\,000 and a wavelength coverage from 3780 to 6910\,\AA{},
with a small gap between 5259 and 5337\,\AA{}.
The data was reduced on La Silla using the HARPS data reduction pipeline.
The normalization of the spectra to the continuum level is described in detail by \citet{Hubrig2013}. 

\subsection{VLT 8m$/$FORS2}
\label{sect:fors}

Twelve spectropolarimetric observations of HD\,57372 have been obtained in service mode using FORS2
at the spectral resolution of about 2\,000
between 2021 November 24 and 2022 April 1.
The data reduction and analysis methods used for $\bz$ measurement were described in detail in
numerous previous papers (e.g., \citealt{Schoeller2017,choj2022})
and will not be repeated here.
As in those papers, the FORS2 data yielded two $\bz$ data sets for the magnetic field measurements:
one using only the hydrogen Balmer series lines ($\bz_{\rm hyd}$)
and another using the entire spectrum from each observation ($\bz_{\rm all}$).

\subsection{TESS lightcurve and rotation period}
\label{sect:tess}

\begin{figure}
\centering
\includegraphics[width=0.95\columnwidth]{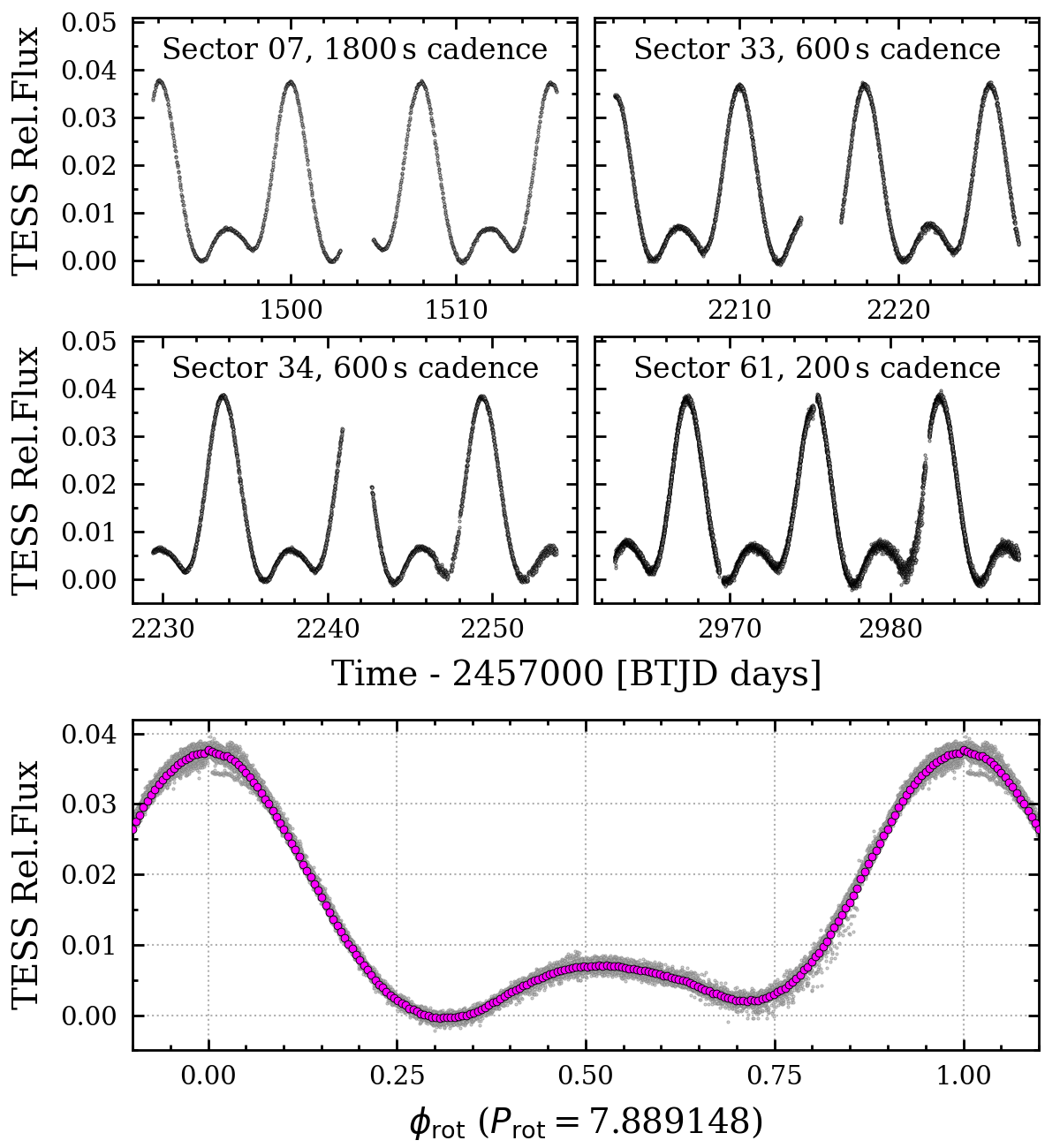}
\caption{
TESS lightcurve of HD\,57372.
The upper panels show the observed data from each of four TESS sectors.
Grey points in the lower panel are all of the data phased by the 7.889148\,d rotation period
and with 2458492.07 being a reference epoch of maximum brightness.
Magenta points show the data binned in increments of $\phi_{\rm rot}=0.005$.
\label{fig:tessfig}
}
\end{figure}

\begin{table*}
\begin{center}
\caption{
Summary of the spectroscopic observations of HD\,57372.
\label{tab:obs}
}
\begin{tabular}{cccccc}
\hline
\hline
UT-Mid  & JD-Mid  & Inst.  & $t_{\rm exp}$ & $S/N$ & $\phi_{\rm rot}$  \\ 
        & 2.4E6+  &        & [s]            &     &   \\ 
\hline
2020-12-29T07:56:09 & 59212.831 &       A & 4899 & \phantom{2}240 & 0.3610 \\ 
2021-01-03T07:40:17 & 59217.820 &       A & 3919 & \phantom{2}259 & 0.9934 \\ 
2021-01-04T07:48:46 & 59218.826 &       A & 3919 & \phantom{2}230 & 0.1209 \\ 
2021-01-06T07:57:01 & 59220.831 &       A & 2939 & \phantom{2}264 & 0.3752 \\ 
2022-04-23T23:48:11 & 59693.492 &       H & 3600 & \phantom{2}115 & 0.2879 \\ 
2022-04-24T23:47:23 & 59694.491 &       H & 3600 & \phantom{2}111 & 0.4146 \\ 
2022-04-25T23:35:26 & 59695.483 &       H & 3600 & \phantom{2}117 & 0.5403 \\ 
2022-04-27T23:30:03 & 59697.479 &       H & 3600 & \phantom{2}116 & 0.7933 \\ 
2022-04-29T00:06:04 & 59698.504 &       H & 3600 & \phantom{2}\phantom{2}95 & 0.9233 \\ 
2022-04-29T23:24:50 & 59699.476 &       H & 3600 & \phantom{2}\phantom{2}92 & 0.0464 \\ 
2022-05-01T00:33:06 & 59700.523 &       H & 3600 & \phantom{2}137 & 0.1792 \\ 
2022-05-02T00:23:28 & 59701.516 &       H & 3600 & \phantom{2}105 & 0.3051 \\ 
2021-11-24T06:53:34 & 59542.787 &       F & $8\times104$ & 2735 & 0.1851 \\ 
2022-01-10T01:28:33 & 59589.561 &       F & $8\times104$ & 2266 & 0.1141 \\ 
2022-01-12T01:26:32 & 59591.560 &       F & $8\times104$ & 2992 & 0.3674 \\ 
2022-01-18T07:18:20 & 59597.804 &       F & $8\times104$ & 2305 & 0.1589 \\ 
2022-02-04T01:50:44 & 59614.577 &       F & $8\times104$ & 2345 & 0.2849 \\ 
2022-02-16T03:02:18 & 59626.627 &       F & $8\times104$ & 2938 & 0.8123 \\ 
2022-03-13T04:45:24 & 59651.698 &       F & $8\times104$ & 2352 & 0.9903 \\ 
2022-03-14T05:00:23 & 59652.709 &       F & $8\times104$ & 2536 & 0.1184 \\ 
2022-03-22T01:31:43 & 59660.564 &       F & $8\times104$ & 1905 & 0.1141 \\ 
2022-03-23T03:47:31 & 59661.658 &       F & $8\times104$ & 2491 & 0.2528 \\ 
2022-03-31T01:13:26 & 59669.551 &       F & $8\times104$ & 2188 & 0.2533 \\ 
2022-04-01T01:11:34 & 59670.550 &       F & $8\times104$ & 2626 & 0.3798 \\ 
\hline
\end{tabular}
\end{center}
Notes:
In the Inst.\ column, A stands for APOGEE, H stands for HARPS, and F stands for FORS2.
HARPS $S/N$ values were estimated at 5500\,{\AA} and APOGEE $S/N$ values were estimated at 16000\,{\AA}.
\end{table*}

HD\,57372 was observed 1086 times in TESS Sector~7 (120\,s cadence) between 2019 January~8 and 2019 February~1,
3485 times in TESS Sector~33 (600\,s cadence) between 2020 December~18 and 2021 January~13,
3474 times in TESS Sector~34 (600\,s cadence) between 2021 January~14 and 2021 February~8,
and 10654 times in TESS Sector~61 (200\,s cadence) between 2023 January~18 and 2023 February~12,
bringing the grand total to just over 100\,days of TESS exposure time spread out over 18,699 individual observations.
We used the \textsc{lightkurve} \citep{Lightkurve2018} code to download the simple aperture photometry,
to which a pixel mask accounting for blending with neighbour stars has been applied.
After removing systematic trends and outliers via the \textsc{TESScut} code \citep{Brasseur2019},
we were left with a total of 18,064 TESS data points.
The lightcurves from each sector are shown in the upper panels of Figure~\ref{fig:tessfig}.

Analysis of the TESS data with the time series software package \textsc{Period04} \citep{Lenz2005}
indicates a primary frequency of $f=0.12675795\pm 2.1\times 10^{-7}$,
which corresponds to a rotation period of $P_{\rm rot}=7.889148\pm 1.3\times 10^{-5}$.
This result agrees well with what has been found in past studies, albeit with far greater precision.
In the bottom panel of Figure~\ref{fig:tessfig}, all of the TESS data has been phase folded
by $P_{\rm rot}$ using 2458492.07 as a JD of maximum brightness.
As is common for Ap/Bp stars
\citep[e.g.][]{Jagelka2019}
due to inhomogeneous elemental abundances and magnetic field strengths across their surfaces,
the lightcurve of HD\,57372 is quite irregular.
Each rotation of HD\,57372 witnesses two local brightness minima at $\phi\sim0.32$ and $\sim0.72$
that enclose a local brightness maximum roughly halfway between at $\phi\sim0.52$.

With rotation phases ($\phi_{\rm rot}$) in hand, Table~\ref{tab:obs}
summarises the spectroscopic observations of HD\,57372, providing Universal Times (UT) at mid-exposure,
Julian dates (JD) at mid-exposure minus 2400000, instruments used,
exposure times, signal-to-noise ratios ($S/N$), and $\phi_{\rm rot}$.

\section{Gaia DR3 parameters and radial velocity measurements}
\label{sect:gaia}

Although HD\,57372 appears to be a single star in the 2MASS and DSS images provided on the HD\,57372 SIMBAD page,
the star has been identified as a visual double star in past studies.
For example, the Tycho Double Star Catalogue \citep{Fabricius2002} reported a component separation of $\sim2${\arcsec},
with the brighter component having magnitudes of $B_{T}=8.15$ and $V_{T}=8.30$
and with the fainter component having $B_{T}=9.05$ and $V_{T}=9.11$.
The third data release of the Gaia mission \citep[Gaia DR3;][]{gaia,gaiadr3}
also reported two sources within 2{\arcsec} of HD\,57372, one with $G=8.289$ and one with $G=9.121$.
The Gaia DR3 parallaxes and proper motions of the brighter star
($\pi=2.412$\,mas, $\mu_{\alpha}=-7.928$\,mas\,yr$^{-1}$, $\mu_{\delta}=4.130$\,mas\,yr$^{-1}$)
are quite similar to those of the fainter star
($\pi=2.378$\,mas, $\mu_{\alpha}=-8.592$\,mas\,yr$^{-1}$, $\mu_{\delta}=4.263$\,mas\,yr$^{-1}$)
such that it is possible the two stars are physically associated. In this case, the orbital period
will of the order of several thousand years.
Spectroscopic parameters were only given for the fainter star.
These included a radial velocity of $v_{\rm r}=24.63$\,km\,s$^{-1}$ as well as an effective temperature of $T_{\rm eff}=13963$\,K that is in
good agreement with the B8 temperature class quoted for HD\,57372 in the \citet{RensonManfroid2009} catalogue of CP stars.

The radial velocity reported by Gaia for the fainter component of HD\,57372 is quite close to the
values we find from the APOGEE and HARPS spectra of the brighter, magnetic component.
For the radial velocity measurements using the APOGEE $H$-band spectra of HD\,57372
we relied on the \ion{Mg}{ii}\,16765\,\AA{}, \ion{Ca}{ii}\,16654\,\AA{},
\ion{Ce}{iii}\,15961\,\AA{}, and \ion{Ce}{iii}\,16133\,\AA{} lines.
These are among the strongest identified lines covered, and their entire profiles could be satisfactorily
fit with single Gaussians during epochs when the magnetic field strength was relatively low
such that the magnetic splitting was not resolved.
In the spectra where magnetic splitting was resolved, we fitted Gaussians to the central components
of the quasi-triplet features of the \ion{Ca}{ii} and \ion{Ce}{iii}, discarding the \ion{Mg}{ii} line
due to its distorted, non-triplet morphology.
Barycentric velocity corrections were applied to the resulting averages of each spectrum,
and the standard deviations of the line-by-line measurements were adopted as error estimates.
This yielded an overall average of $v_{\rm r}=21.88\pm2.80$\,km\,s$^{-1}$ across the four APOGEE spectra,
with individual $v_{\rm r}$ ranging from 21.30--22.87\,km\,s$^{-1}$.
From this, we have no reason to suspect that HD\,57372 is a member of a binary system, or at least not of a short period binary.

In the HARPS optical spectra, $v_{\rm r}$ was measured similarly
using \ion{Cr}{ii} and \ion{Fe}{ii} lines.
The lines were selected on the basis of relatively short wavelengths ($<4800$\,\AA{})
and relatively low effective Land\'{e} factors ($g_{\rm eff}<0.7$), to minimise the effects of magnetic splitting.
This yielded an overall average of $v_{\rm r}=22.45\pm0.86$\,km\,s$^{-1}$ across the eight HARPS spectra,
with individual $v_{\rm r}$ ranging from 21.63--23.51\,km\,s$^{-1}$.
The surprisingly large $v_{\rm r}$ scatter here seems to be caused by an inhomogeneous
chemical element distribution on the stellar surface typically observed in Ap/Bp stars and
a complicated magnetic field geometry rather than by binarity. However, our attempts to phase the
  individual $v_{\rm r}$ values with
the rotation period did not show any conclusive correlation.

Although the Gaia data does not indicate the fainter component of HD\,57372 being variable, the brighter component is classified
as a photometric variable
with a 7.888\,d period. Considering that this is in good agreement with the rotation period of the magnetic component of HD\,57372
indicated by our data, and also considering the aforementioned radial velocity agreement, we proceed here under the assumption that the
aforementioned Gaia DR3 spectroscopic parameters
(including also the surface gravity of $\log g=4.399$
and the stellar radius of $R_{*}=1.88\,R_{\odot}$) in fact pertain to the Bp star HD\,57372 that is the focus of this paper.
Assuming that the TESS 7.889\,d period is in fact the rotation period of HD\,57372,
then the $R_{*}=1.88\,R_{\odot}$ would imply a rotational velocity at the equator $v_{eq}=12$\,km\,s$^{-1}$.

The relatively small angular separation of the two objects may lead to contamination
of the primary spectrum by the secondary. The APOGEE fibers subtend 2${\arcsec}$ on the sky,
whereas the HARPS fibers have a diameter of 1${\arcsec}$, and the used FORS2 slit was 0.4${\arcsec}$. Therefore, for
the APOGEE observations we expect that some light of the secondary was captured.
For HARPS and FORS2, contamination will be rather small. However, we do not see
any evidence for a second star in our data.
Further, as we show in the following sections, all observables vary
according to a rotation period that can be
independently measured from photometry, Zeeman splitting variability, line strength
variability, and longitudinal field variability, confirming that they are tied to one
single star. All results presented in our paper are typical for a single magnetic
star, albeit an extreme one.

\section{Magnetic field measurements}
\label{sect:mfield}

\subsection{$H$-band variability of the magnetic field modulus}
\label{sect:magvarH}

\begin{figure*}
\centering
\includegraphics[width=0.95\textwidth]{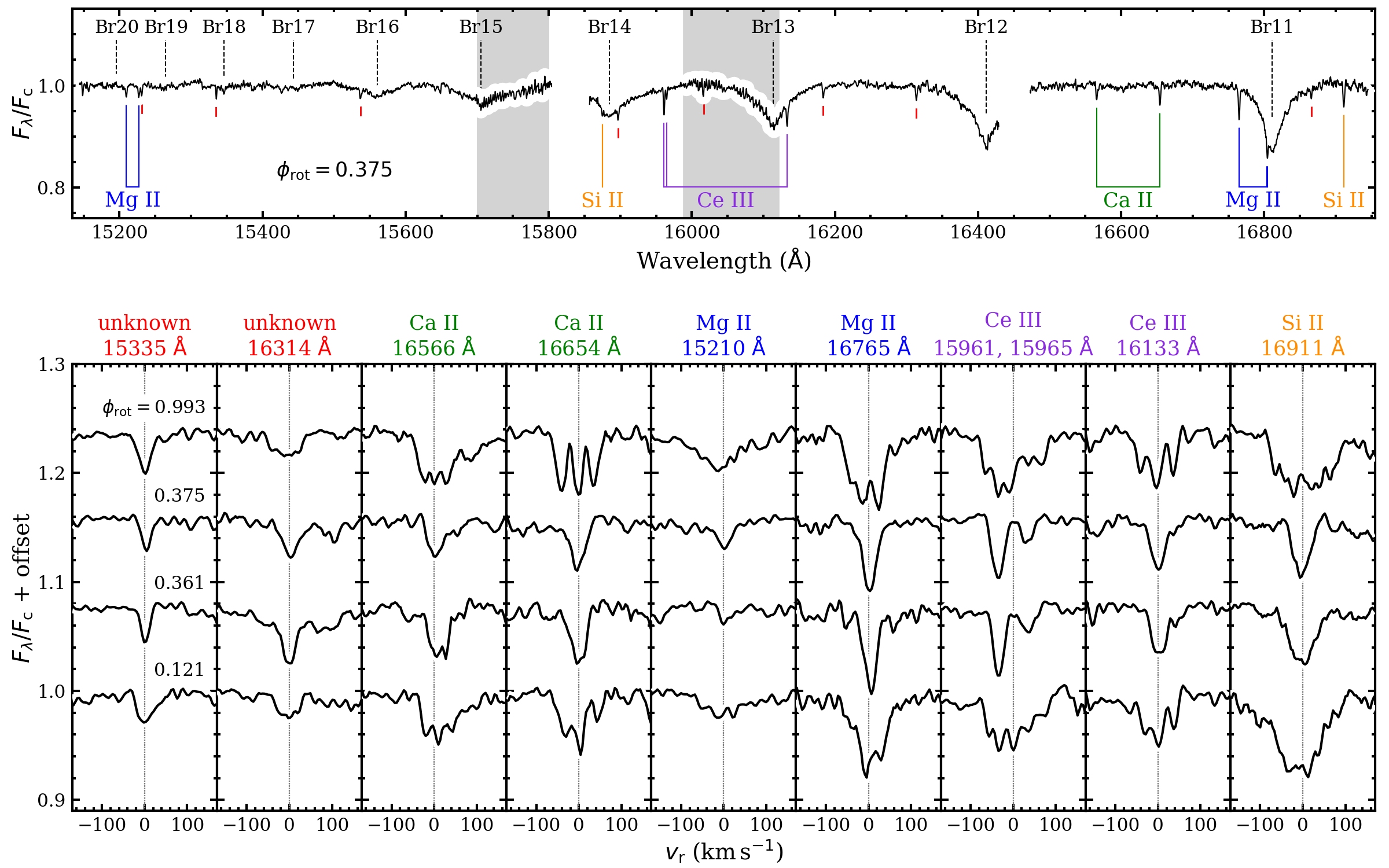}
\caption{
  \emph{Upper panel:} line identifications in the APOGEE spectrum of HD\,57372 at the rotational phase $\phi_{\rm rot}=0.375$.
 The hydrogen Brackett series lines are labelled along the top, while metal lines are labelled along the bottom.
Unidentified lines are indicated by small red ticks below the spectrum.
\emph{Lower panel:} variability of individual lines as a function of $\phi_{\rm rot}$.
\label{fig:apvar}
}
\end{figure*}

We present in the upper panel of Figure~\ref{fig:apvar} line identifications in the $H$-band spectra of HD\,57372.
As is typical for late-Bp stars, the hydrogen Brackett series are the strongest features in the spectra,
while the metal line content is dominated by the rare earth ion \ion{Ce}{iii} along with singly-ionized elements
such as \ion{Mg}{ii}, \ion{Si}{ii}, and \ion{Ca}{ii}.
We also show in the spectrum of HD\,57372 most of the unidentified lines that were previously
discussed by \citet{choj2019,choj2020,choj2023}.
These features have no likely counterparts in the existing atomic data.

The lower panel of Figure~\ref{fig:apvar} highlights the remarkable variability of spectral lines over the course
of one week.
Whereas the $\phi_{\rm rot}=0.361$ and 0.375 spectra show hardly any evidence of magnetic splitting,
most of the lines are clearly split by a strong magnetic field during the $\phi_{\rm rot}=0.121$ and 0.993 spectra.
The best examples of this are \ion{Ca}{ii}\,16654\,\AA{} and \ion{Ce}{iii}\,16133\,\AA{},
both of which have quasi-triplet Zeeman patterns. 

Atomic transitions in which the upper and lower energy levels both have total angular momentum
  quantum numbers equal to 0.5 (i.e. $J_{1}=J_{2}=0.5$) but only one of the levels is split (i.e. with a non-zero Land\'{e} factor,
  $g_{1}=0$, $g_{2}>0$ or $g_{1}>0$, $g_{2}=0$) are known as Zeeman doublets.
 In absence of the partial Paschen-Back effect, which should not have too strong impact at the modest field strengths in question,
 this combination of $J$ and $g$ values leads to a Zeeman pattern consisting of two $\pi$ components, each of
  which overlaps in wavelength or velocity space with a corresponding $\sigma$ component. In these cases, the mean magnetic field modulus
  can be estimated via $\bm$ (G)$=\Delta\lambda/(k\,g\,\lambda_{0}^{2})$ \citep{mathys1997} where $g$ is the Land\'{e} factor of the split
  level, $k$ is a constant equal to 4.671$\times$10$^{-13}$\,kG$^{-1}$\,\AA{}$^{-1}$, $\lambda_{0}$ is the rest wavelength of the transition
  in units of \AA{}, and $\Delta\lambda$ is the separation of the Zeeman doublet components in units of \AA{}.
For a pure Zeeman triplet, $g$ is either the Land\'{e} factor
of the split level of the transition, if the other level is unsplit, or if the transition is between two split levels that
have the same Land\'{e} factor, the value of the latter.

No such transitions are present in the $H$-band spectra of HD\,57372, so in order to estimate $\bm$ from the Zeeman quasi-triplet features
  that are present, we replace $g$ with the effective Land\'{e} factors \citep[$g_{\rm eff}$; see equation 4 of][]{1990A&A...236..527M} of
  the transitions and $\Delta\lambda$ with half of the separation of the outer triplet components. This makes it straightforward to
  estimate $\bm$ from the $\phi_{\rm rot}=0.121$ and 0.993 APOGEE spectra of HD\,57372. The most cleanly split lines in these spectra
  are \ion{Ca}{ii}\,16654\,\AA{} and \ion{Ce}{iii}\,16133\,\AA{}, and from them we find $\bm\sim14.5$\,kG at $\phi_{\rm rot}=0.121$
  and $\sim15.3$\,kG at $\phi_{\rm rot}=0.993$.

\begin{figure}
\centering
\includegraphics[width=0.95\columnwidth]{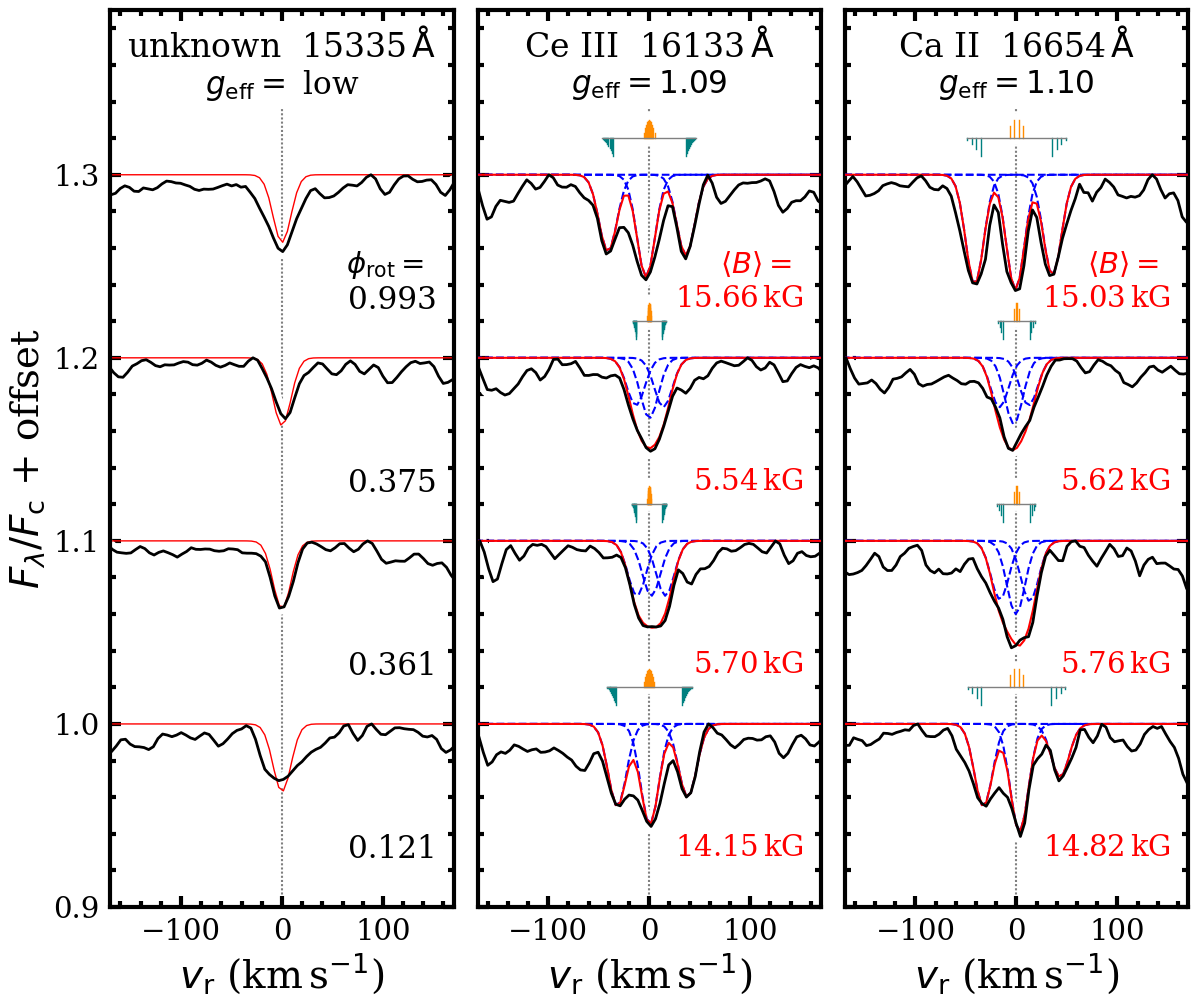}
\caption{
Mean magnetic field modulus estimates from the \ion{Ce}{iii}\,16133\,\AA{} and \ion{Ca}{ii}\,16654\,\AA{} lines
in the APOGEE $H$-band spectra.
The relatively magnetically insensitive unknown line at 15335\,\AA{} is shown in order to demonstrate
that it is well fit by a single Gaussian with FWHM$=21$\,km\,s$^{-1}$ during the low-$\bm$ epochs.
For \ion{Ce}{iii}\,16133\,\AA{} and \ion{Ca}{ii}\,16654\,\AA{},
the triple Gaussians each have the same FWHM$=21$\,km\,s$^{-1}$.
The associated $\bm$ estimates are given below each line profile,
but it is important to note that the low-$\bm$ values are highly uncertain
due to the magnetic splitting being unresolved at APOGEE's resolution
 \label{fig:hzsplit}
}
\end{figure}

The question now is, how low is $\bm$ during the $\phi_{\rm rot}=0.361$
and $\phi_{\rm rot}=0.375$ spectra? Figure~\ref{fig:hzsplit} represents our attempt to answer this question.
We have yet to see resolved magnetic splitting in the unknown line at 15335\,\AA{}
in the APOGEE spectra of any star, meaning that it has a quite low $g_{\rm eff}$
and thus is relatively insensitive to the magnetic field.
The 15335\,\AA{} line is by far the narrowest line in the APOGEE spectra of HD\,57372,
and it has an observed minimum FWHM of $\sim21$\,km\,s$^{-1}$ in the $\phi_{\rm rot}=0.361$ spectrum.
To demonstrate this, a Gaussian of this FWHM has been plotted over the 15335\,\AA{} line profiles.

In the \ion{Ca}{ii}\,16654\,\AA{} and \ion{Ce}{iii}\,16133\,\AA{} panels of Figure~\ref{fig:hzsplit},
we have assumed that $\sim21$\,km\,s$^{-1}$ is roughly the intrinsic FWHM in the absence of a magnetic field,
and we have also approximated the Zeeman patterns as simple ensembles of three Gaussians with fixed FWHM$=21$\,km\,s$^{-1}$.
The positions and depths of the individual Gaussians have then been adjusted manually in order to arrive at the displayed results.
Although the quoted $\bm$ values are highly uncertain during the $\phi_{\rm rot}=0.361$ and 0.365 spectra,
the implication is that $\bm$ has dropped below 6\,kG and hence that $\bm$ changes by nearly 10\,kG over the course of
each rotation cycle.

It is worth noting the high degree of asymmetry of the line profiles in the $\phi_{\rm rot}=0.121$ spectrum.
This is atypical of the APOGEE-observed Ap/Bp stars that exhibit RMSL,
and is likely due to the rotational Doppler shifts of the contributions to the observed,
disc-integrated lines coming from different parts of the stellar
disc.

\subsection{Optical variability of the magnetic field modulus}
\label{sect:magvarV}

\begin{figure}
\centering
\includegraphics[width=0.95\columnwidth]{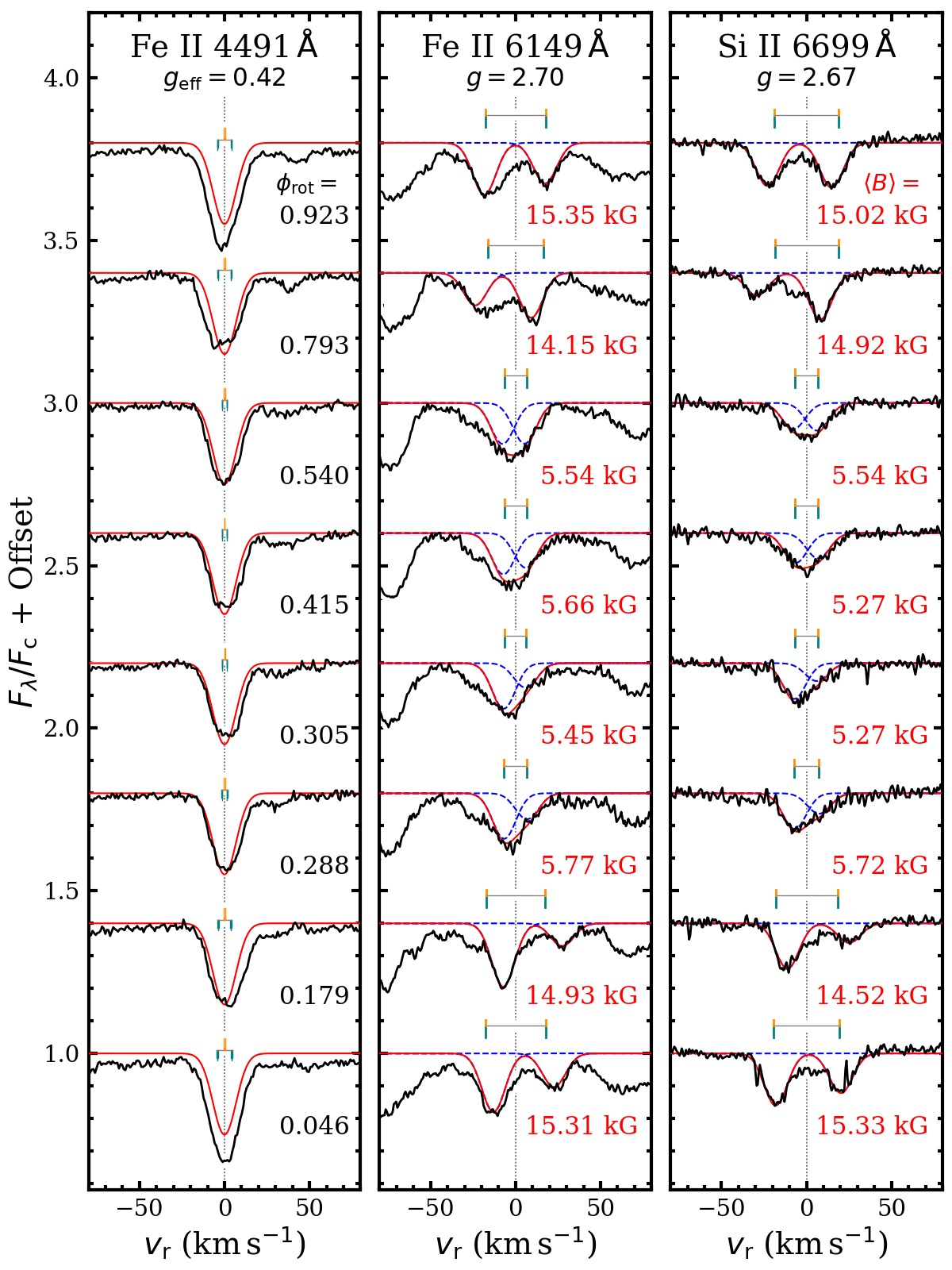}
\caption{
Similar to Figure~\ref{fig:hzsplit} but for the HARPS optical spectra.
As in that figure, the best available magnetically insensitive line (\ion{Fe}{ii}\,4491\,\AA{})
is also shown, and the displayed Gaussians for the lines \ion{Fe}{ii}\,6149\,\AA{} and
\ion{Si}{ii}\,6699\,\AA{} have fixed FWHM$=$16\,km\,s$^{-1}$.
This value is based on the \ion{Si}{ii}\,6699\,\AA{}, where it gives the best fit to the magnetically split doublet
components at e.g. rotation phases 0.046, 0.793, and 0.923. The Gaussian used to fit the line
\ion{Fe}{ii}\,4491\,\AA{} has a fixed FWHM$=14$\,km\,s$^{-1}$. Its depth 
has been fixed to match the observed depth in the $\phi_{\rm rot}=0.540$ spectrum.
\label{fig:ozsplit}
}
\end{figure}

Before discussing the rich line content of the optical HARPS spectra of HD\,57372, we repeated with the HARPS spectra the exercise 
detailed in the previous subsection for estimating $\bm$ though in this case making use of Zeeman doublet features. Similar 
to the APOGEE spectra, RMSL are observed in numerous lines in half of the HARPS spectra while $\bm$ was too weak for any RMSL to be 
observed in the other half. The middle and righthand panels of Figure~\ref{fig:ozsplit} show the two best available Zeeman doublet 
features -- \ion{Fe}{ii}\,6149\,\AA{} ($g=2.70$) and \ion{Si}{ii}\,6699\,\AA{} ($g=2.67$). Several other doublets 
such as \ion{Ti}{ii}\,4315\,\AA{} ($g=2.63$) and \ion{Fe}{ii}\,4385\,\AA{} ($g=2.68$) are also covered by the HARPS 
spectra, but we focus on the longer wavelength lines since blending is far more of a concern in the 4300\,\AA{} region and because 
the Zeeman splitting is larger for a fixed $g$ or $g_{\rm eff}$ as wavelength increases. 

Also shown in Figure~\ref{fig:ozsplit} is \ion{Fe}{ii}\,4491\,\AA{},
which is the best available of a magnetically insensitive ($g_{\rm eff}=0.43$) line
that shows no resolved splitting in any of the HD\,57372 spectra.
Whereas lines with $g_{\rm eff}=0$ (e.g. \ion{Fe}{i}\,5434\,\AA{})
are often present in the optical spectra of Ap stars, HD\,57372 is sufficiently hot that \ion{Fe}{i} lines are
very weak or not present.
During the low-$\bm$ HARPS spectra, the \ion{Fe}{ii}\,4491\,\AA{} has an observed FWHM of roughly 16\,km\,s$^{-1}$.
Taking into account unresolved splitting that will slightly broaden the line,
the Gaussian used to fit this line shown in Figure~\ref{fig:ozsplit} has a fixed FWHM$=14$\,km\,s$^{-1}$. 
 
As can be seen in this figure, the \ion{Si}{ii}\,6699\,\AA{} line,
provides a satisfactory fit to the magnetically split doublet components
during the low- and high-$\phi_{\rm rot}$ phases where the splitting is resolved.
In the case of \ion{Fe}{ii}\,6149\,\AA{} however,
which is the most frequently used $\bm$ diagnostic since it is present
and relatively strong for the majority of Ap stars \citep{mathys2017},
the splitting is during all epochs contaminated by one or more unknown features
such that a simple double Gaussian composite is insufficient for a satisfactory fit.
The presence of  the \ion{Si}{ii}\,6699\,\AA{} line, on the other hand, requires quite severe Si enhancement,
and not only is this the case for HD\,57372, the line is quite isolated with no likely contaminants in the vicinity.
Our placement of the \ion{Fe}{ii}\,6149\,\AA{} double Gaussians in Figure~\ref{fig:ozsplit}
was therefore guided by what is shown for \ion{Si}{ii}\,6699\,\AA{}, which we regard as the ground truth in this case.

It is important to note that no algorithmic fitting occurred in the creation of Figures~\ref{fig:hzsplit} and \ref{fig:ozsplit}.
Rather, the positions and depths of the Gaussians were all adjusted manually in order to achieve qualitative best fits.
These plots are useful for guiding the eye, but they do not provide formal errors.
The level of uncertainty is clearly quite high between $\phi_{\rm rot}=0.2$--0.7 given
that magnetic splitting was unresolved in the associated spectra. 

Even in the spectra where the splitting was resolved, the behaviour of the splitting
is quite different with respect to most RMSL stars for three reasons.
First, in all of the HARPS spectra of HD\,57372 that exhibit RMSL,
there is a high degree of asymmetry in the depths of the low- and high-velocity Zeeman components.
At low $\phi_{\rm rot}$, the high velocity components are shallower than the low velocity components,
and the depths of the shallower components diminish as $\phi_{\rm rot}$ increases.
Second, the positions of the components appear to drift towards higher velocity as $\phi_{\rm rot}$ increase from $\phi_{\rm rot}=0$.
These behaviours then appear to reverse at some intermediate $\phi_{\rm rot}$,
such that the mirror image occurs as $\phi_{\rm rot}$ approaches unity.
Third, the observed line profiles of the unblended, pure Zeeman doublet \ion{Si}{ii}\,6699\,\AA{} line
cannot be accounted for with a simple two Gaussian composite during any of the $\phi_{\rm rot}$ extrema spectra.
On the other hand, there is no reason to expect the shapes of the split components to be Gaussian, as even for very slowly rotating
stars, they depend on the distribution of the magnetic field strength over the visible stellar disk.
Given the non-negligible value of the projected equatorial velocity, the shape difference between the blue and red components of
the \ion{Fe}{ii} and \ion{Si}{ii} doublets
in HD\,57372 reflects the different combinations of Zeeman and Doppler effects on different parts of the stellar surface.
The similar behaviour of the
\ion{Fe}{ii} and \ion{Si}{ii} doublet lines considered here strengthens this interpretation, even though part of the component difference in
\ion{Fe}{ii}  must also result from partial Paschen-Back effect.

\subsection{Longitudinal magnetic field}
\label{sect:Bz}

\begin{figure*}
\includegraphics[width=\textwidth]{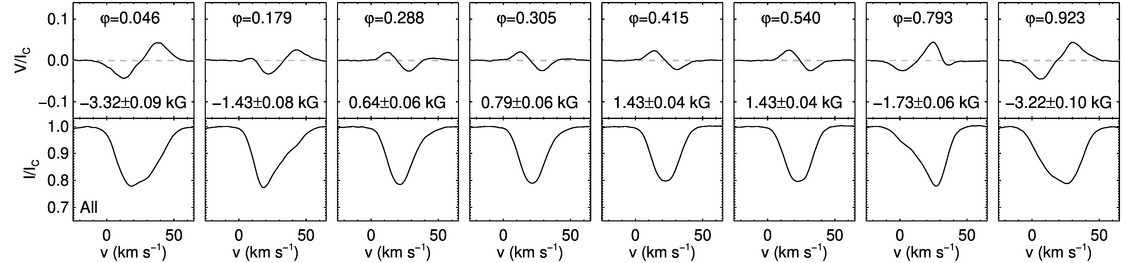}
\caption{LSD Stokes~$I$ (bottom), and Stokes~$V$ (top) calculated using HARPS spectropolarimetric observations
  of HD\,57372 recorded on different rotation phases.
  The employed line mask includes all identified spectral lines that appear to be unblended or minimally
blendedin.  The rotational phases and the measured longitudinal 
magnetic field strengths are presented on the top. The dashed 
lines in the Stokes~$V$ panels indicate the mean uncertainties, which for the presented line mask are of 
the order of the thickness of these lines.
\label{fig:IVN_all}}
\end{figure*}

\begin{table*}
\begin{center}
\caption{
Radial velocity and magnetic field measurements.
\label{tab:btable}
}
\begin{tabular}{ccr@{$\pm$}lcccr@{$\pm$}lr@{$\pm$}l}
\hline
\hline
Inst. & 
$\phi_{\rm rot}$ & 
\multicolumn{2}{c}{$v_{\rm r}$} &
 $\bm$ &
 $\bm$ &
 $\bm$ &
\multicolumn{2}{c}{$\bz$} &
\multicolumn{2}{c}{$\bz$} \\ 
 &
 &
\multicolumn{2}{c}{} &
\ion{Si}{ii}\,6699 &
\ion{Ce}{iii}\,16133 &
\ion{Ca}{ii}\,16654 &
\multicolumn{2}{c}{all} &
\multicolumn{2}{c}{hydrogen} \\ 
 &
 &
\multicolumn{2}{c}{[km\,s$^{-1}]$} &
[kG] &
[kG] &
[kG] &
\multicolumn{2}{c}{[kG]} &
\multicolumn{2}{c}{[kG]} \\ 
\hline
H & 0.046 & 23.14 & 0.74           & 15.33 &       &       & $-3.32$ & 0.09 & \multicolumn{2}{c}{  } \\ 
F & 0.114 & \multicolumn{2}{c}{  } &       &       &       & $-4.40$ & 0.08 & $-3.66$ & 0.10 \\ 
F & 0.114 & \multicolumn{2}{c}{  } &       &       &       & $-4.31$ & 0.08 & $-3.63$ & 0.09 \\ 
F & 0.118 & \multicolumn{2}{c}{  } &       &       &       & $-4.26$ & 0.08 & $-3.59$ & 0.08 \\ 
A & 0.121 & 21.71 & 3.22           &       & 14.15 & 14.82 & \multicolumn{2}{c}{  }      & \multicolumn{2}{c}{  } \\ 
F & 0.159 & \multicolumn{2}{c}{  } &       &       &       & $-3.00$ & 0.07 & $-2.61$ & 0.10 \\ 
H & 0.179 & 22.49 & 1.04           & 14.47 &       &       & $-1.43$ & 0.08 & \multicolumn{2}{c}{  } \\ 
F & 0.185 & \multicolumn{2}{c}{  } &       &       &       & $-2.12$ & 0.14 & $-1.66$ & 0.18 \\ 
F & 0.253 & \multicolumn{2}{c}{  } &       &       &       & $-0.08$ & 0.04 & $-0.03$ & 0.07 \\ 
F & 0.253 & \multicolumn{2}{c}{  } &       &       &       & $-0.07$ & 0.04 & $-0.03$ & 0.07 \\ 
F & 0.285 & \multicolumn{2}{c}{  } &       &       &       & 0.74    & 0.04 & 0.72 & 0.07 \\ 
H & 0.288 & 21.81 & 0.55           & 5.72  &       &       & 0.64    & 0.06 & \multicolumn{2}{c}{  } \\ 
H & 0.305 & 21.63 & 0.55           & 5.27  &       &       & 0.79    & 0.06 & \multicolumn{2}{c}{  } \\ 
A & 0.361 & 21.30 & 2.67           &       & 5.70  & 5.76  & \multicolumn{2}{c}{  }      & \multicolumn{2}{c}{  } \\ 
F & 0.367 & \multicolumn{2}{c}{  } &       &       &       & 1.72    & 0.09 & 1.65 & 0.10 \\ 
A & 0.375 & 21.64 & 2.74           &       & 5.54  & 5.62  & \multicolumn{2}{c}{  }      & \multicolumn{2}{c}{  } \\ 
F & 0.380 & \multicolumn{2}{c}{  } &       &       &       & 1.70    & 0.04 & 1.65 & 0.05 \\ 
H & 0.415 & 21.70 & 0.46           & 5.23  &       &       & 1.43    & 0.04 & \multicolumn{2}{c}{  } \\ 
H & 0.540 & 22.31 & 0.76           & 5.54  &       &       & 1.43    & 0.04 & \multicolumn{2}{c}{  } \\ 
H & 0.793 & 22.98 & 1.33           & 14.92 &       &       & $-1.73$ & 0.06 & \multicolumn{2}{c}{  } \\ 
F & 0.812 & \multicolumn{2}{c}{  } &       &       &       & $-3.29$ & 0.07 & $-2.74$ & 0.09 \\ 
H & 0.923 & 23.51 & 1.06           & 15.02 &       &       & $-3.22$ & 0.10 & \multicolumn{2}{c}{  } \\ 
F & 0.990 & \multicolumn{2}{c}{  } &       &       &       & $-6.05$ & 0.09 & $-5.25$ & 0.10 \\ 
A & 0.993 & 22.87 & 2.53           &       & 15.66 & 15.03 & \multicolumn{2}{c}{  }      & \multicolumn{2}{c}{  } \\ 
\hline
\end{tabular}
\end{center}
Notes:
Instrument column meanings are as in Table~\ref{tab:obs}.
In the $\bz$ met/all column, HARPS values pertain to $\bz$ as measured from apparently unblended metal lines while FORS2 values pertain to $\bz$ as measured from the full spectra.
\end{table*}

\subsubsection{HARPS observations}
\label{sect:Bz_HARPS}

As we have done in our previous studies using HARPSpol data \citep[see, e.g.][]{Jarvinen2020},
we employed the least-squares deconvolution (LSD) technique following the description given by \citet{Donati1997}
in order to increase the accuracy of the mean longitudinal magnetic field ($\bz$) determination on different
rotation phases.
The parameters of the lines used to calculate the LSD profiles were taken
from the Vienna Atomic Line Database \citep[VALD3;][]{Kupka2011}.
Only lines that appear to be unblended or minimally
blended in the Stokes~$I$ spectra were included in the line mask.
The resulting profiles are scaled according to the line strength and 
sensitivity to the magnetic field. The final line mask contains 87 lines and includes
Si, Ti, Cr, Fe, and Sr lines, also lines belonging to the 
rare-earth element (REE group such as Nd, Dy, and Er). The LSD 
profiles calculated with this line mask are shown in Fig.~\ref{fig:IVN_all}. Since the diagnostic null
polarization spectrum calculated according to the definition of \citet{Donati1997} 
is featureless and perfectly flat, it is not presented in this figure.
The mean longitudinal magnetic field is usually determined by computing the first-order 
moment of the LSD Stokes~$V$ profile according to
\citet{Mathys1989}.

Further,  as numerous studies of Ap/Bp stars in the past have revealed 
that different elements typically show different abundance distributions across the stellar surface,
we carried out LSD magnetic field measurements using line masks for each element separately.
These previous studies also indicated a kind of symmetry between the topology of the magnetic field
and the element distribution. Thus, the structure of the magnetic field
can be potentially studied by measurements of the magnetic field using
spectral lines of each element separately.
In Fig.~\ref{afig:IVNiron} we present the measurements 
of the longitudinal magnetic field using line masks containing
\ion{Ti}{ii}, \ion{Cr}{ii}, \ion{Fe}{ii}, and \ion{Sr}{ii} lines,
whereas Fig.~\ref{afig:IVNree}  shows the results for twice ionised 
REE elements \ion{Dy}{iii}, \ion{Er}{iii}, and \ion{Nd}{iii}. 
Additionally, we present in Fig.~\ref{afig:IVNSi} magnetic field measurements
carried out separately for \ion{Si}{ii} and 
\ion{Si}{iii} lines to to check the presence of an anomalies related to the
vertical abundance stratification frequently observed in Ap/Bp stars (e.g. \citealt{Hubrig2018}).
The measured mean longitudinal magnetic field strengths for all line masks are presented in Table~\ref{tab:bz}.

The presence of a magnetic field in the LSD profile is evaluated according to \citet{Donati1992},
who define that a Zeeman profile with a false alarm probability (FAP) $\leq 10^{-5}$ is considered
as a definite detection, $10^{-5} <$ FAP $\leq 10^{-3}$ as a marginal detection,
and FAP $> 10^{-3}$ as a non-detection.
All used line masks for all eight epochs give definite detections.

The general behaviour of the mean 
longitudinal magnetic field strength calculated for all masks is rather similar: the field is positive throughout 
observed rotation phases 0.288--0.540 and negative in the phases 0.793--0.179. For the field
calculated using the line mask with all 87 spectral lines the strongest
negative field ($-3.32\pm0.09$\,kG) is measured around the phase 0, whereas the strongest 
positive field ($1.43\pm0.04$\,kG) is detected around the phase 0.5.
The observed single change of the field polarity hints at a prevailing dipolar field structure.
The LSD 
Stokes~I profiles presented in Fig.~\ref{fig:IVN_all} appear
asymmetric when the field has negative polarity, indicating the presence of an inhomogeneous element
surface distribution and are almost symmetric when the field shows positive polarity.

The longitudinal magnetic field strengths measured using \ion{Sr}{ii} and the iron-peak elements
\ion{Ti}{ii}, \ion{Cr}{ii}, and \ion{Fe}{ii} show a  very 
similar behaviour. The strongest field of negative polarity is measured 
around the phase 0: \ion{Ti}{ii} lines give the strongest field of 
$-3.68\pm0.18$\,kG followed by \ion{Cr}{ii} and \ion{Fe}{ii} whereas using
 \ion{Sr}{ii} lines the field is clearly weaker with the strength of $-2.51\pm0.43$\,kG. 
\ion{Ti}{ii} lines show the strongest positive field of 
$1.82\pm0.08$\,kG at the phase 0.540. The positive magnetic longitudinal 
magnetic field measured using \ion{Sr}{ii} LSD profiles shows
maximum field strength at phase the phase 0.288. The presence of asymmetric Stokes~$I$ profiles
in the phases with the negative field indicate surface inhomogeneous abundance distribution of these elements.

Interestingly, the longitudinal magnetic fields of negative polarity measured using different REE line masks  
(Fig.~\ref{afig:IVNree}) appear to be much stronger than fields measured using other elements.
This can be explained that, similar to several previous studies of Ap and Bp stars, these elements
are distributed on the stellar surface closer to the location of the magnetic pole of
negative polarity (e.g., \citealt{Hubrig2018,Jarvinen2020}). 
The measurements using \ion{Dy}{iii} lines account for $-5.56\pm0.40$\,kG and 
those for \ion{Nd}{iii} and \ion{Er}{iii} are around $-4$\,kG. On the other hand, the
positive longitudinal magnetic field strengths for REEs are comparable 
with those obtained for iron peak elements. The Stokes~$I$ profiles exhibit a distinct structure 
in all rotation phases, indicating a surface inhomogeneous abundance distribution.

We do not find any indication for the presence of a vertical stratification
of Si: the line profiles 
using both \ion{Si}{ii} and \ion{Si}{iii} lines are very similar. Also the 
measured longitudinal fields do not differ much apart from the negative field extrema
with $\bz$=$-3.25\pm0.15$\,kG measure for \ion{Si}{ii} lines and $\bz$=$-2.28\pm0.09$\,kG for
\ion{Si}{iii} lines.
Otherwise magnetic fields measured using both 
line lists follow the common pattern being most negative around the phase 0 
and most positive around phase 0.5.

\subsubsection{FORS2 observations}
\label{sect:Bz_FORS}

Twelve FORS2 spectropolarimetric observations
of HD\,57372 were obtained in service mode over a few months from 2021 to 2022. Unfortunately, a few observations cover
similar rotation phases, e.g. around the phases 0.1, 0.25, and 0.37, implying larger gaps in the rotation phase
coverage between 0.38 and 0.81, and 0.81 and 0.99.
The measurements of the mean longitudinal magnetic field were carried
out using procedures presented in prior work (e.g., \citealt{Hubrig2004a,Hubrig2004b,Schoeller2017}).
The obtained longitudinal magnetic field strengths using for the measurements only the hydrogen Balmer series lines
($\bz_{\rm hyd}$) and using the entire spectrum ($\bz_{\rm all}$) present clear rotational modulation,
which is expected for a large-scale organised dipole field structure and
show change of polarity, with the strongest mean longitudinal magnetic field of negative
polarity $\bz_{\rm all}$= $-$6.05$\pm$0.09\,kG in the rotation phase 0.990 and the strongest field of positive
polarity $\bz_{\rm all}$= 1.72$\pm$0.09\,kG in the rotation phase 0.367.

\begin{figure}
\centering
\includegraphics[width=0.95\columnwidth]{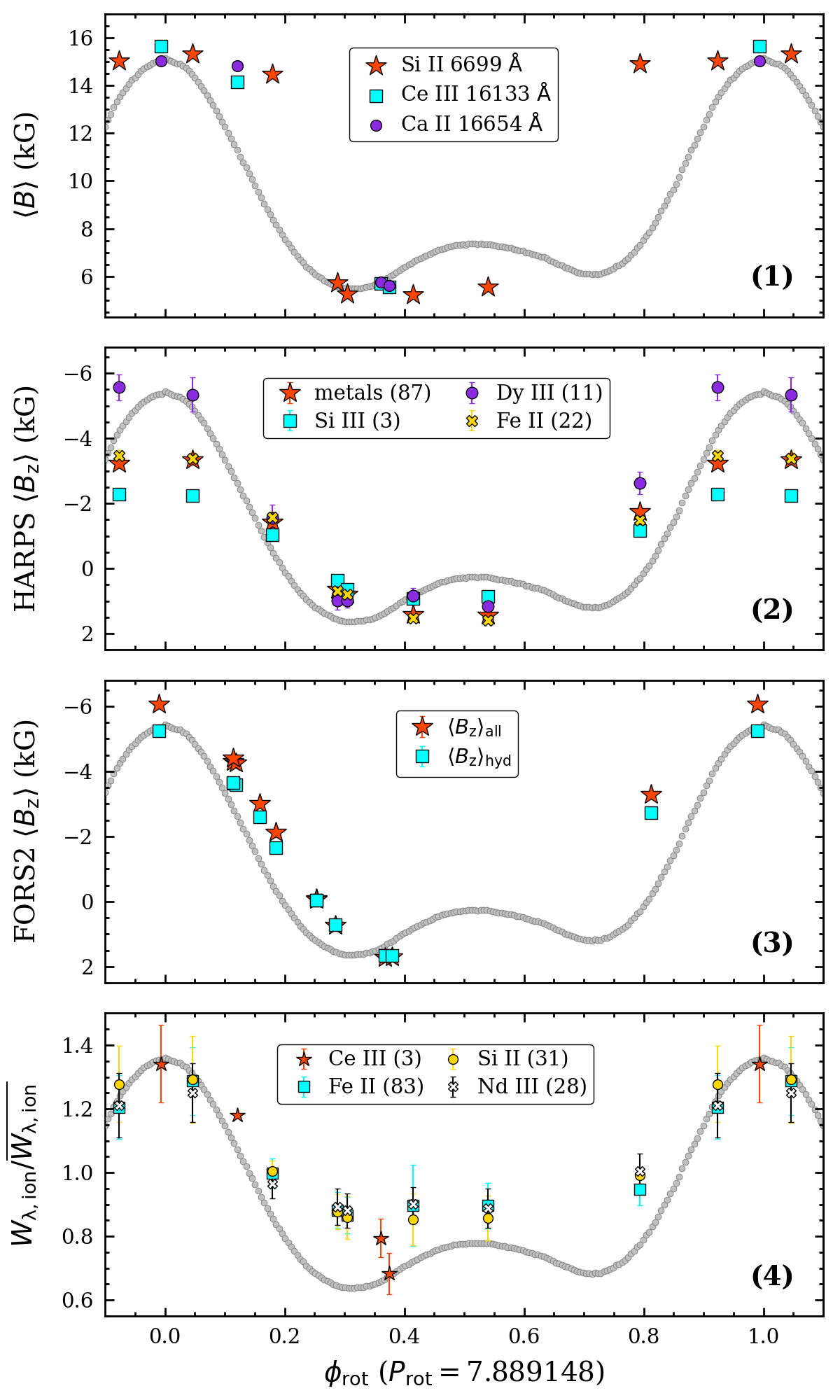}
\caption{
Variability of HD\,57372 as a function of rotation phase.
The phase-binned TESS lightcurve (magenta points in Figure~\ref{fig:tessfig}) is shown as grey points in each panel.
The ordinate scaling of the lightcurve is similar to that of the bottom panel of Figure~\ref{fig:tessfig}, but the associated
  y-axis tick labels have been omitted to emphasize the units of the scatter points.
Panel~(1) shows the $\bm$ estimates from the \ion{Si}{ii}\,6699\,\AA{} line
in the HARPS spectra and from the \ion{Ce}{iii}\,16133\,\AA{} and \ion{Ca}{ii}\,16654\,\AA{} lines in the APOGEE spectra.
Panel~(2) shows the HARPS $\bz$ measurements, including those obtained from numerous metal lines,
from three \ion{Si}{iii} lines, from 11 \ion{Dy}{iii} lines, and from 22 \ion{Fe}{ii} lines.
These ions were selected simply to demonstrate the larger degree of scatter in $\bz$ when $\bm$ is high.
Panel~(3) shows the $\bz$ measurements from the FORS2 data, including those from the full spectra and those from the hydrogen lines.
Panel~(4) shows the equivalent width variations for four example ions, with the numbers of lines
averaged for each ion given in parentheses.
\label{fig:magvar}
}
\end{figure}

In Table~\ref{tab:btable} we present, along with the radial velocity measurements
reported in Sect.~\ref{sect:gaia},
our magnetic field measurements
using all NIR and optical observations at our disposal, acquired with different instruments.
Photometric, spectroscopic, and magnetic variability of HD\,57372 as a function of rotation phase
is presented in Fig.~\ref{fig:magvar}. All measurements show a good correlation with the TESS light curve exhibiting
stronger (in absolute values) mean magnetic field moduli, mean longitudinal magnetic fields, and
line profile intensities of the studied elements in rotational phases around the light curve maximum.
Remarkably, equivalent widths change by up to 50\% of their mean values
over the course of a rotation period.

\section{Atmospheric parameters and abundances}
\label{sect:stellar_pars}

Our visual inspection of the available high-resolution spectra of HD\,57372 suggests that
this star, similar to other magnetic Ap/Bp stars, is extreme in abundances of several elements.
In order to estimate the stellar parameters and surface abundances of HD\,57372,
we created a small grid of synthetic spectra using the SYNTHE program with
ATLAS9 model atmospheres \citep{kurucz2005} and the VALD atomic linelist.
To minimise the impact of magnetic broadening, we focussed our analysis
on the average of the four HARPS spectra taken during intermediate rotational phases
when no magnetic splitting is observed ($\phi_{\rm rot}=0.288$, 0.305, 0.415, 0.540).
The rotational velocity was fixed to $v\sin i=14$\,km\,s$^{-1}$ based
on the width of \ion{Fe}{ii}~4491\,\AA{}, which as previously noted is one
of the least magnetically sensitive lines present in the spectra.
The microturbulent velocity was fixed to 2\,km\,s$^{-1}$.

\begin{table}
\begin{center}
\caption{
Abundance estimates from the mean low-$\bm$ HARPS spectrum of HD\,57372.
\label{tab:abu}
}
\begin{tabular}{lrr@{$\pm$}rr}
\hline
\hline
\multicolumn{1}{c}{Atom/Ion} &
\multicolumn{1}{c}{$Z$} &
\multicolumn{2}{c}{[X$/$H] (dex)} &
\multicolumn{1}{c}{$N$} \\ 
\hline
\ion{He}{i}   & 2.00  & $-$0.05 & 0.02 & 1 \\ 
\ion{C}{ii}   & 6.01  & 0.81    & 0.05 & 3 \\ 
\ion{N}{ii}   & 7.01  & 1.41    & 0.38 & 6 \\ 
\ion{O}{i}    & 8.00  & 1.00    & 0.14 & 3 \\ 
\ion{O}{ii}   & 8.01  & 1.73    & 0.11 & 5 \\ 
\ion{Ne}{i}   & 10.00 & 0.35    & 0.19 & 4 \\ 
\ion{Mg}{ii}  & 12.01 & 0.96    & 0.20 & 10 \\ 
\ion{Al}{ii}  & 13.01 & 0.73    & 0.16 & 5 \\ 
\ion{Al}{iii} & 13.02 & 1.52    & 0.16 & 5 \\ 
\ion{Si}{ii}  & 14.01 & 1.83    & 0.28 & 36 \\ 
\ion{Si}{iii} & 14.02 & 2.63    & 0.05 & 4 \\ 
\ion{P}{ii}   & 15.01 & 0.80    & 0.04 & 2 \\ 
\ion{S}{ii}   & 16.01 & 1.42    & 0.24 & 33 \\ 
\ion{Cl}{ii}  & 17.01 & 3.14    & 0.35 & 27 \\ 
\ion{Ca}{ii}  & 20.01 & 1.92    & 0.25 & 9 \\ 
\ion{Sc}{ii}  & 21.01 & 2.13    & 0.12 & 5 \\ 
\ion{Ti}{ii}  & 22.01 & 2.16    & 0.29 & 38 \\ 
\ion{V}{ii}   & 23.01 & 0.86    & 0.06 & 2 \\ 
\ion{Cr}{ii}  & 24.01 & 1.71    & 0.24 & 34 \\ 
\ion{Mn}{ii}  & 25.01 & 0.76    & 0.00 & 2 \\ 
\ion{Fe}{ii}  & 26.01 & 1.61    & 0.18 & 64 \\ 
\ion{Fe}{iii} & 26.02 & 2.09    & 0.22 & 9 \\ 
\ion{Co}{ii}  & 27.01 & 1.92    & 0.06 & 2 \\ 
\ion{Ni}{ii}  & 28.01 & 0.47    & 0.34 & 2 \\ 
\ion{Sr}{ii}  & 38.01 & 2.94    & 0.04 & 2 \\ 
\ion{Ce}{iii} & 58.02 & 3.84    & 0.08 & 2 \\ 
\ion{Pr}{iii} & 59.02 & 3.87    & 0.14 & 3 \\ 
\ion{Nd}{iii} & 60.02 & 4.09    & 0.20 & 28 \\ 
\ion{Tb}{iii} & 65.02 & 3.94    & 0.36 & 7 \\ 
\ion{Dy}{iii} & 66.02 & 3.96    & 0.35 & 19 \\ 
\ion{Ho}{iii} & 67.02 & 3.59    & 0.05 & 3 \\ 
\ion{Er}{iii} & 68.02 & 4.13    & 0.20 & 8 \\ 
\hline
\end{tabular}
\end{center}
Notes:
The abundances are given relative to the solar abundances presented by \citet{asplund2021}. 
The number of spectral lines used for the abundance determination is presented
in the last column.
\end{table}

\begin{figure*}
\centering
\includegraphics[width=0.95\textwidth]{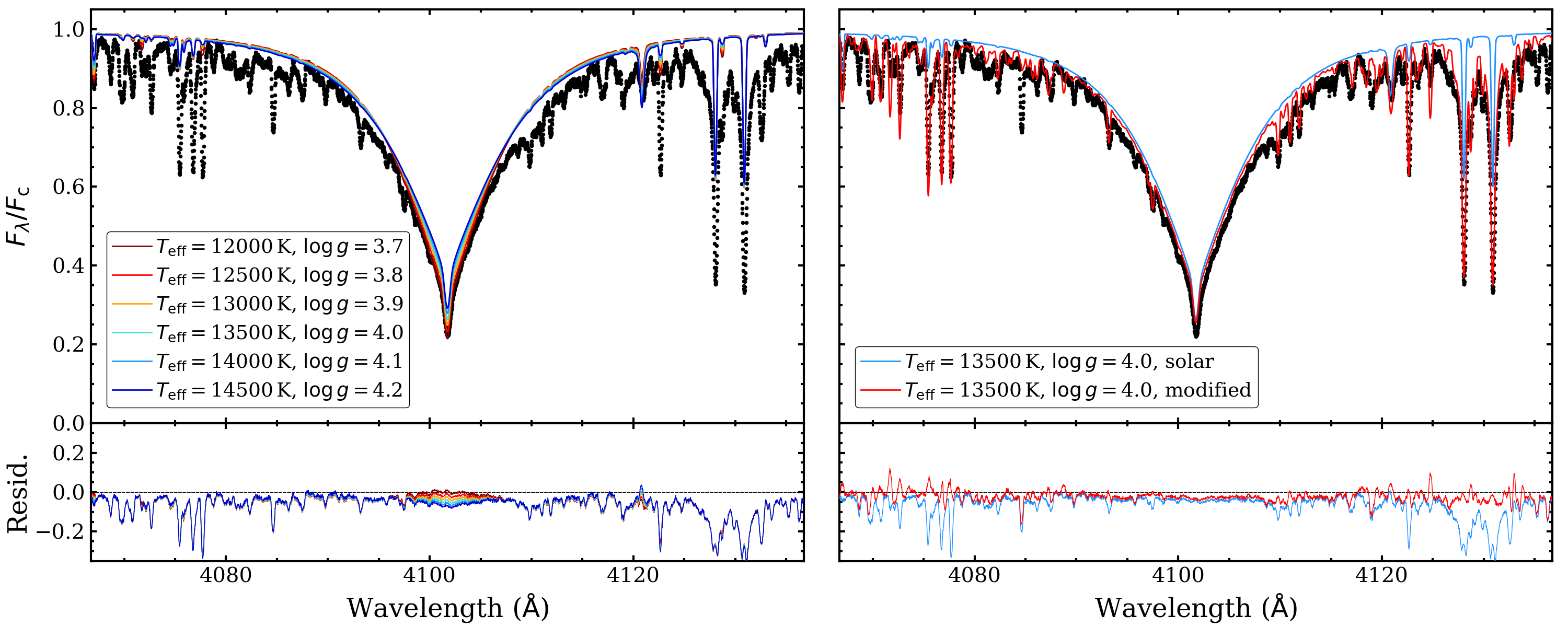}
\caption{
\emph{Left:} Comparison of the observed H$\delta$ line (black dots) from the mean low-$\bm$ spectrum of HD\,57372
to solar abundance synthetic spectra (coloured curves) covering a range of $T_{\rm eff}$ and $\log g$.
\emph{Right:} Similar to the left panel, but now with the blue curve showing the solar abundance synthetic spectrum
with the adopted $T_{\rm eff}=13500$\,K and $\log g=4.0$, and with the red curve being the same
but with abundances adjusted as quoted in Table~\ref{tab:abu}.
\label{fig:hsynth}
}
\end{figure*}

The effective temperature ($T_{\rm eff}$) and surface gravity ($\log g$)
were assessed based on the Balmer series lines with some consideration given to the behaviour of metal lines,
for example the weakness of the \ion{Fe}{i} lines and the strength of the \ion{Fe}{iii} lines.
Whereas neutral helium is a good $T_{\rm eff}$ diagnostic for normal late-B stars, it is usually found to be depleted on the surfaces of late-Bp stars like HD\,57372.
A total of 78 synthetic spectra were created, with $T_{\rm eff}$ ranging from 11500$-$15000\,K
in steps of 500\,K and with $\log g$ ranging from 3.7$-$4.5 in steps of 0.1 dex.
The left panel of Figure~\ref{fig:hsynth} shows some of the synthetic spectra
that provide a reasonable match to the H$\delta$ line of HD\,57372. 

\begin{figure}
\centering
\includegraphics[width=0.95\columnwidth]{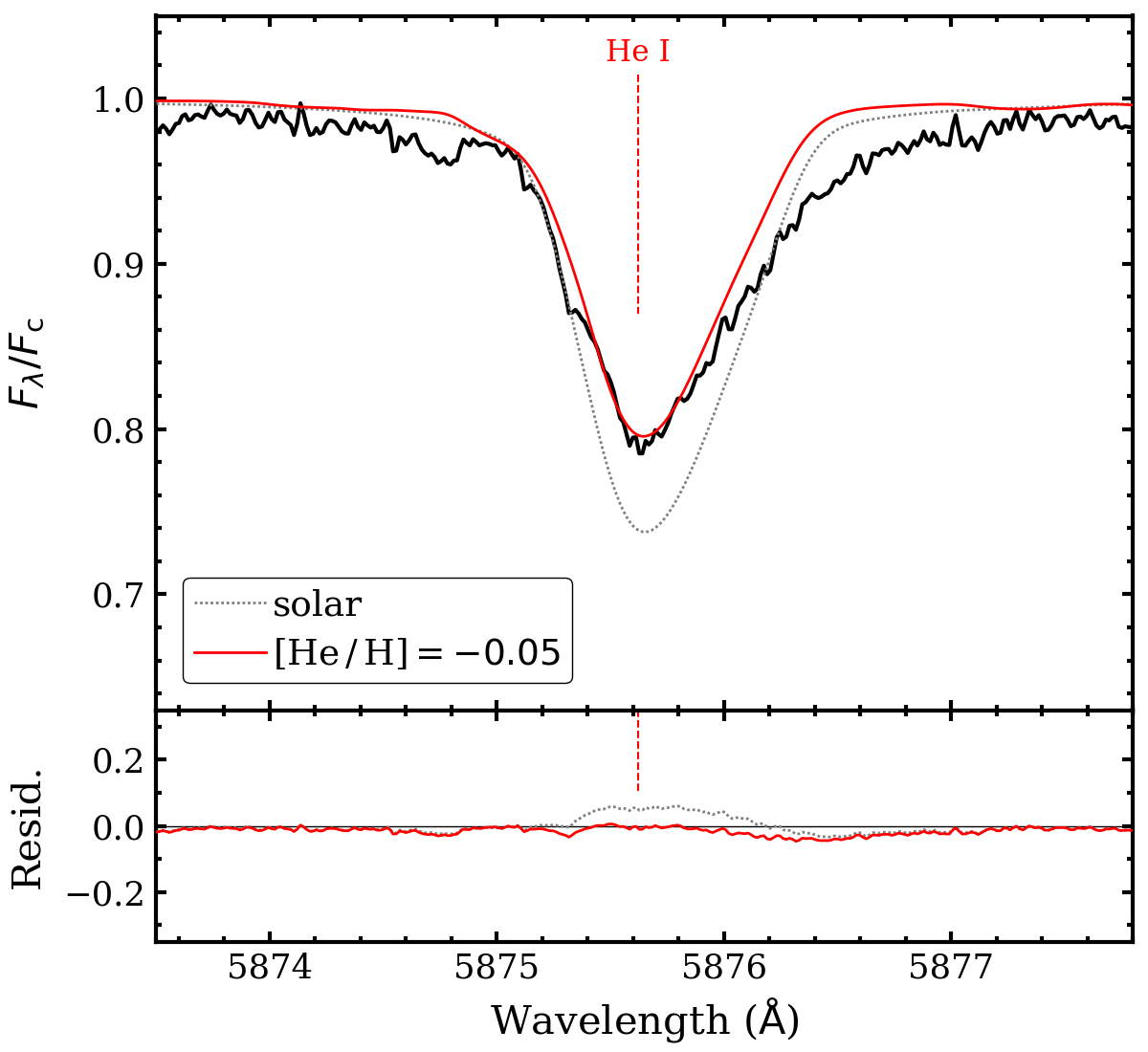}
\caption{
Estimate of the helium abundance from the \ion{He}{i} 5875\,\AA{} line
for the case of $T_{\rm eff}=13500$\,K and $\log g=4.0$.
The dotted line is a synthetic spectrum with solar helium abundance
while the solid red line has a decreased helium abundance.
Observed minus synthetic residuals are shown in the low panel.
\label{fig:hesynth}
}
\end{figure}

\begin{figure*}
\centering
\includegraphics[width=0.95\textwidth]{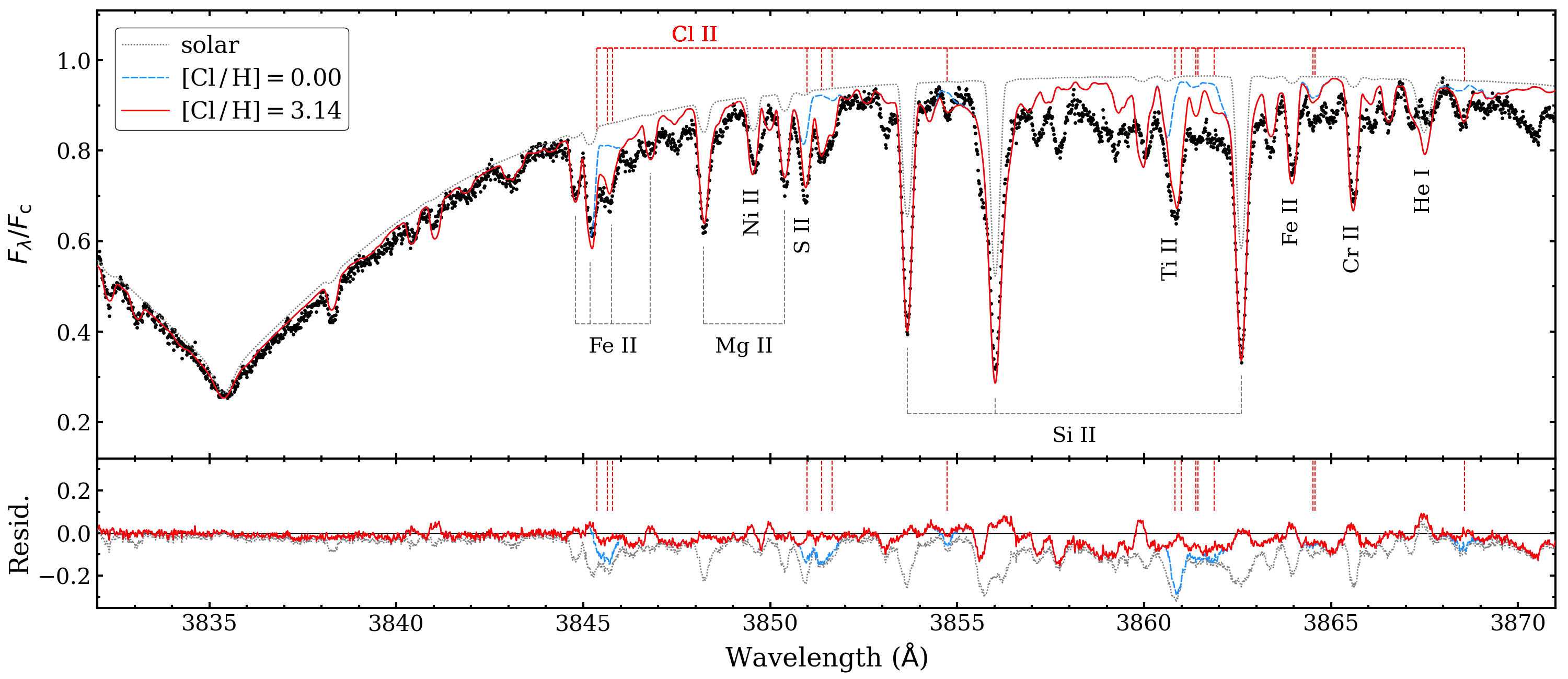}
\caption{
Estimate of the ionised chlorine abundance.
The observed spectrum is shown as black dots.
As indicated in the legend, three synthetic spectra are shown.
The dotted grey curve has solar abundances across the board and the dashed blue curve
has all abundances except for Cl adjusted to the values given in Table~\ref{tab:abu}.
 The solid red curve is then the same as blue except for [Cl$^{+}/$H] having been increased by 3.14 dex.
\label{fig:clsynth}
}
\end{figure*}

Although the model spectrum with $T_{\rm eff}=12500$\,K and $\log g=3.8$
provides a good fit to the cores of the Balmer series lines, it is important to note a few following aspects.
First, the optical spectra of HD\,57372 contain a plethora of lines with no reasonable counterparts in the available atomic data.
It is therefore reasonable to expect the depths of the hydrogen lines to be blanketed to some extent.
Second, while \ion{Si}{i} and \ion{Fe}{i} are all but absent from the observed spectra,
lines from \ion{Si}{ii}, \ion{Si}{iii}, \ion{Fe}{ii}, and \ion{Fe}{iii} are plentiful.
Adoption of $T_{\rm eff}=12500$\,K results not only in neutral Si and Fe lines that are far too strong,
but also in large discrepancies between the singly-ionized versus doubly-ionized abundances of Si and Fe. 

As for the hotter model spectra, those with $T_{\rm eff}\geq14000$\,K are disfavoured
due to the poor fits to the cores of the Balmer lines.
We thus conclude that the $T_{\rm eff}$ of HD\,57372 is likely in the range 12500--14500\,K
and that the $\log g$ is likely in the range 3.8--4.2, and proceed to estimate abundances
for the case of $T_{\rm eff}=13500$\,K and $\log g=4.0$.
Due to the aforementioned prevalence of unknown lines in the spectra of HD\,57372
as well as the inevitable effects of magnetic broadening, even during the low-$\bm$ epochs,
we used the equivalent width ($W_{\lambda}$) matching technique in order to derive abundances. 

The vast majority of identifiable lines in the observed spectra can be attributed to \ion{Fe}{ii},
so we began by estimating the iron abundance based on 78 lines that appeared to be minimally blended.
For each line, the continuum normalization was adjusted in 300\,km\,s$^{-1}$ windows
centred on the rest wavelengths and $W_{\lambda}$ was then measured via direct flux integration
in fixed 20\,km\,s$^{-1}$ windows centred on the rest wavelengths.
The same procedure was then applied to synthetic spectra, with [Fe/H] being adjusted
until the synthetic $W_{\lambda}$ matched the observed $W_{\lambda}$.
Obvious outliers (mostly due to blending with unknown lines) were then rejected,
and the line-by-line abundances were averaged.
A total of 64 lines yielded [Fe$^{+}/$H]\,$=1.610\pm0.182$.
From visual inspection, it was clear that the doubly-ionized iron abundance
did not agree with that of singly-ionized iron, so the \ion{Fe}{iii} abundance
was estimated independently in the same manner using an initial list of nine lines
that appeared to be free of blending with \ion{Fe}{ii}.
This yielded [Fe$^{++}/$H]\,$=2.089\pm0.222$ from a total of nine lines.
All subsequent abundance estimates were carried with fixed [Fe$/$H]\,$=1.610$.

The next strongest and most numerous contributions to the observed spectra beyond iron
include \ion{Ti}{ii}, \ion{Si}{ii}, \ion{S}{ii}, and \ion{Cr}{ii}.
We proceeded over these elements iteratively by first estimating the \ion{Ti}{ii} abundance
and keeping it constant for subsequent abundance estimates,
then estimating the \ion{Si}{ii} abundance and keeping it constant for subsequent abundance estimates,
and so on for a total of 31 ions that we are confident contribute to the observed spectra.
The only exceptions to this general process were oxygen, silicon, and aluminuim,
for which both singly-ionized and doubly-ionized lines are present.
Similar to the case of iron, the abundances of different ionization stages were estimated independently for these elements.
In the right panel of Figure~\ref{fig:hsynth} we show a good fit of the observed H$\delta$ line by the 
atmosphere model with abundances adjusted as quoted in Table~\ref{tab:abu}.

With some noteworthy exceptions,
the trend in the abundance results presented in Table~\ref{tab:abu} is generally similar to those of the majority of Ap/Bp
stars whose abundances have been previously estimated (e.g. \citealt{Ghazaryan2018}).
As is demonstrated in Figure~\ref{fig:hesynth}, helium is clearly depleted with respect to solar.
This is a common feature of Ap/Bp stars photospheres, as is the strong enhancement
of silicon, iron peak elements, and strontium, and the very strong enhancement of rare earth elements.
However, one element certainly sticks out due to its remarkable overabundance:
we detect a strong chlorine enhancement of 3.14\,dex above solar. It appears to be 
stronger than ever reported for any Ap/Bp star in the past, albeit with just six stars in the literature for which 
chlorine abundances have been reported.
Our estimate of the ionised chlorine abundance using three different synthetic spectra is
presented in Figure~\ref{fig:clsynth}.

\section{Discussion}
\label{sect:discussion}

\begin{figure}
\centering
\includegraphics[width=0.95\columnwidth]{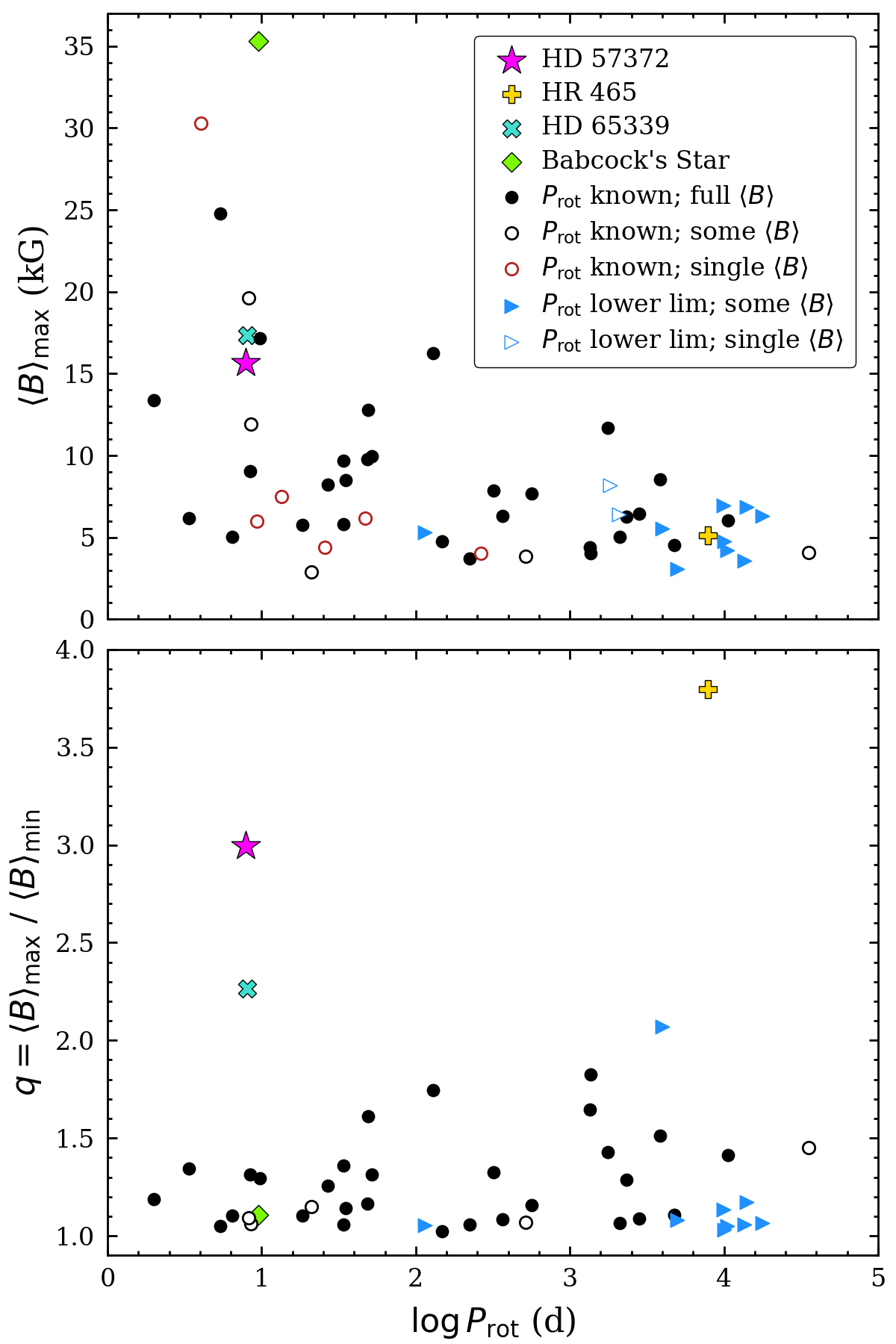}
\caption{
Comparison of HD\,57372 to other stars known to exhibit resolved, magnetically split lines in the optical.
\emph{Upper panel:} the maximum or only available $\bm$ estimate as a function of $P_{\rm rot}$.
\emph{Lower panel:} the ratio of $\bm_{\rm max}$ over $\bm_{\rm min}$ as a function of $P_{\rm rot}$.
In both panels, some of the more exceptional stars are indicated by their own symbol and colour.
\label{fig:compare}
}
\end{figure}

HD\,57372 possesses a very strong magnetic field, with an exceptionally
large amplitude of variation of the magnetic field modulus of about 10\,kG over the rotation period of 7.889\,d.
The maximal strength of $\bm$=15.66\,kG is measured
using the Zeeman triplet \ion{Ce}{iii}\,16133\,\AA{}  and the lowest $\bm$=5.23\,kG is estimated using the Zeeman doublet
\ion{Si}{ii}\,6699\,\AA{}.
All observable quantities are found to vary in phase.
This includes the longitudinal magnetic field $\bz$, which varies in FORS2 spectra from roughly $-6$\,kG up to $+1.7$\,kG,
as well as the metal line strengths, whose equivalent widths change by up to 50\% of their mean values
over the course of a rotation period.
In the high-resolution spectra recorded in visual and NIR spectra, we detect 
significant variability of the line profiles over the rotation cycle. We also observe in Stokes $I$ spectra 
significant distortion of the resolved components of the Zeeman doublets \ion{Si}{ii}\,6699\,\AA{} and
\ion{Fe}{ii}\,6149\,\AA{}. Such distortions are most likely due to the combination of different amounts of
Doppler and Zeeman effects on different parts of the stellar surface covered by chemical spots.

The exceptional value of 3 of the ratio between the maximum and the minimum
of the mean magnetic field modulus that is observed in this Bp star is indicative of a very unusual
geometry of its magnetic field.
In Fig.~\ref{fig:compare}  we compare HD\,57372 to other Ap/Bp stars known to exhibit
resolved, magnetically split lines in the optical. Most of the measurements were provided by \citet{mathys1997}, \citet{elkin2012}, \citet[][and references therein]{mathys2017}, and \citet{giarrusso2022}. 
In the top panel of this figure  we present the maximum or only available $\bm$ estimates as a function of $P_{\rm rot}$.
Only a handful of other stars show stronger $\bm$ than HD\,57372.
The distribution of the ratio $q$ of the maximum to the minimum of the mean magnetic field
modulus measured in HD\,57372 is presented in the bottom of Fig.~\ref{fig:compare} along
with the ratio of $\bm_{\rm max}$ over $\bm_{\rm min}$ for other Ap/Bp stars as a function of $P_{\rm rot}$.
Among the presented stars, only HR\,465 shows a stronger value for the ratio $q$,
of the order of 3.8 \citep{mathys1997,giarrusso2022}, with the second strongest ratio of about 2.3 determined for
HD\,65339 \citep{mathys1997}.
It is of interest that the magnetic field of the Ap star HD\,65339 with the rotation period of 8.03\,d
(\citealt{mathys2017}, and references therein)
shows a complex structure with the best fit
of the $\bm$ variation curve using the superposition of a cosine wave
and of its first harmonic (e.g. \citealt{mathys2017}, and references therein).
As for the Bp star HR\,465 with a very long rotation period of about 22.7\,yr (\citealt{mathys2017}, and
references therein),
\citet{giarrusso2022} conclude that its magnetic field structure should show some significant departure from a
centred dipole. 
Inspecting figure 4 in \citet{giarrusso2022}, we note that the shapes of the $\bm$ phase curves of both stars,
HR\,465 and HD\,57372,
show similarly broad, fairly flat minima and sharper-peaked maxima.

Our longitudinal magnetic field measurements show just one change of polarity, suggesting that the
field topology of HD\,57372 is predominantly dipolar.
For Ap/Bp stars, it is quite common to see variations of the strength of their magnetic field.
These variations can easily be described by the oblique rotator model,
which was introduced by \citet{Stibbs1950}.
The stellar magnetic field is rotating with the star
and one can see different aspects of it.
The oblique rotator assumes a dipolar magnetic field tilted with the obliquity angle $\beta$
with respect to the rotation axis, which itself is inclined with respect to the line-of-sight by the angle $i$.
In the cases that $i$ is zero, one is looking at the rotation pole
and no modulation of the magnetic field with rotation can be seen.
The same is true in cases of $\beta=0$, i.e.\ the magnetic field axis is aligned with the rotation axis.

Using this model, the inclination angle $i$
between the rotation axis of the star and the line of sight is usually constrained if the rotation period
and the stellar radius $R$ are known and from the measurements of $v \sin i$.
In Sect.~\ref{sect:gaia} we reported $v_{eq}=12$\,km\,s$^{-1}$ using Gaia DR3 determinations $\log g=4.399$
and the stellar radius of $R_{*}=1.88\,R_{\odot}$) and assuming the TESS phometric period of 7.889\,d.  
On the other hand, we adopted somewhat different stellar parameters, $T_{\rm eff}=13500$\,K and $\log g=4.0$,
in Sect.~\ref{sect:stellar_pars} to achieve a better match to the H$\delta$ line of HD\,57372.
The lower values for $\log g$ and $T_{\rm eff}$ imply a slightly larger stellar radius and therefore
a larger equator velocity $v_{eq}$,
which, however, can not be greater than 14\,km\,s$^{-1}$ estimated in Sect.~\ref{sect:magvarV} using the magnetically
weakly sensitive \ion{Fe}{ii}\,4491\,\AA{} line with the relatively low effective Land\'{e} factor $g_{\rm eff}=0.43$.
The implication of the finding that the values for equator velocity and the projected rotation velocity $v \sin i$ are nearly the same
is that the inclination angle $i$ between the rotation axis of the star and the line of sight is close to 90$^\circ$.
However, for this angle, we would expect equal field strength for both negative and positive poles, whereas our magnetic
field measurements show a  much stronger field of negative polarity.

In the following we assume the inclination angle $i=85\pm5^\circ$ to test whether it is still possible to
describe the magnetic field structure of HD\,57372 by a simple dipolar field geometry. 
The general description for the strength of the observed longitudinal magnetic field for a simple centred dipole
was presented by \citet{Preston1967}:

\begin{equation}
\left< B_{\rm z} \right> = \frac{1}{20} \, \frac{15+u}{3-u} \, B_d (\cos \beta \cos i + \sin \beta \sin i \cos 2\pi t / P)
\label{eq:diagn.oblique_mod}
\end{equation}

\noindent
with
$\left< B_{\rm z} \right>$ the effective magnetic field,
$\beta$ the angle between the rotation axis and the magnetic axis,
$i$ the angle between the rotation axis and the line of sight,
$P$ the rotation period of the star,
$u$ the limb-darkening coefficient and
$B_d$ the strength of the dipolar magnetic field.

The relative amplitude of variation of the longitudinal field
  is usually characterised by the parameter $r$ representing the ratio between $B_{z,{\rm min}}$ and $B_{z,{\rm max}}$.
    In our case with $B_{z,{\rm max}}=-6.0\pm0.1$\,kG and $B_{z,{\rm min}}=1.7\pm0.1$\,kG measured in FORS2 spectra using the entire spectra
  we calculate $r=-3.53\pm0.22$.
Using the formula

\begin{equation}
  r  =  \frac{B_{z,{\rm min}}}{B_{z,{\rm max}}}
  =  \frac{\cos \beta \cos i - \sin \beta \sin i}{\cos \beta \cos i + \sin \beta \sin i}
\label{eq:diagn.oblique_r}
\end{equation}

we calculate a very small
obliquity angle $\beta=8.90\pm8.81^\circ$. With the limb-darkening coefficient of 0.4
\citep{Claret2019} we obtain an incredibly strong and implausible polar magnetic field strength of $B_{d}=-84\pm52$\,kG.
It is not clear whether the exceptional structure of the magnetic field in HD\,57372 is related to the unusually high
field strength or to the presence of significant departures of its magnetic field from the simple
centred dipole magnetic field geometry.

On the other hand, given the poor phase coverage,
in particular between the rotation phases 0.2 and 0.8 where the field modulus becomes smaller and
the longitudinal field measurements indicate positive polarity,
our assumption of sinusoidal field variability may be wrong. Obviously, additional high-resolution
spectrossopic and polarimetric observations with a more dense coverage of the rotation cycle are necessary
to better constrain the phase curves for the mean field modulus and the mean longitudinal field that
can be in fact anharmonical with noticeable deviations from a sinusoid. In our observations,
the $\bm$  maximum coincides with the negative longitudinal magnetic field $\bz_{\rm all}$ extrema
measured in HARPS and FORS2 spectra, while the $\bm$ minimum is found close in phase with positive longitudinal
magnetic fields.  However, in case of a centred dipole where
both poles come alternatively into view as the star rotates, we should also
expect the modulus variation curve to appear as a perfect double wave,
that is, we would see a sinusoid with twice the rotation frequency of the star. 
From the simultaneous consideration of the phase curves of
variation of the longitudinal field and of the mean field
modulus based on a more complete observational dataset, it will become possible to better constrain the
magnetic field structure (e.g., dipole offset
along its axis or superposition of dipole, quadrupole, and octupole components). 

\begin{acknowledgements}
We thank the referee G. Mathys for constructive comments that helped us to improve the paper.
  Funding for the Sloan Digital Sky Survey V has been provided by the Alfred P. Sloan Foundation, the Heising-Simons Foundation, 
the National Science Foundation, and the Participating Institutions. SDSS acknowledges support and resources from the Center 
for High-Performance Computing at the University of Utah. SDSS telescopes are located at Apache Point Observatory, funded by 
the Astrophysical Research Consortium and operated by New Mexico State University, and at Las Campanas Observatory, operated 
by the Carnegie Institution for Science. The SDSS web site is \url{www.sdss.org}.

SDSS is managed by the Astrophysical Research Consortium for the Participating Institutions of the SDSS Collaboration, including 
Caltech, The Carnegie Institution for Science, Chilean National Time Allocation Committee (CNTAC) ratified researchers, 
The Flatiron Institute, the Gotham Participation Group, Harvard University, Heidelberg University, The Johns Hopkins University, 
L’Ecole polytechnique f\'{e}d\'{e}rale de Lausanne (EPFL), Leibniz-Institut f\"{u}r Astrophysik Potsdam (AIP), Max-Planck-Institut 
f\"{u}r Astronomie (MPIA Heidelberg), Max-Planck-Institut f\"{u}r Extraterrestrische Physik (MPE), Nanjing University, National 
Astronomical Observatories of China (NAOC), New Mexico State University, The Ohio State University, Pennsylvania State University, 
Smithsonian Astrophysical Observatory, Space Telescope Science Institute (STScI), the Stellar Astrophysics Participation Group, 
Universidad Nacional Aut\'{o}noma de M\'{e}xico, University of Arizona, University of Colorado Boulder, University of Illinois at 
Urbana-Champaign, University of Toronto, University of Utah, University of Virginia, Yale University, and Yunnan University.

Based on observations made with ESO Telescope at the La Silla Observatory under programme IDs
0108.D-0205(B) and 0109.C-0265(A).

This work has made use of data from the European Space Agency (ESA) mission {\it Gaia}
(\url{https://www.cosmos.esa.int/gaia}), processed by the
{\it Gaia} Data Processing and Analysis Consortium (DPAC, \url{https://www.cosmos.esa.int/web/gaia/dpac/consortium}).
Funding for the DPAC has been provided by national institutions,
in particular the institutions participating in the {\it Gaia} Multilateral Agreement.

This work has made use of the VALD, operated at Uppsala University,
the Institute of Astronomy RAS in Moscow, and the University of Vienna.
\end{acknowledgements}

%
   \bibliographystyle{aa} 
   \bibliography{drew_hd57} 
%






%

\begin{appendix}
\section{Longitudinal magnetic field measurements using HARPS spectropolarimetric observations}
\label{sect:profiles}

The measured mean longitudinal magnetic field strengths for all line masks are presented in Table~\ref{tab:bz}.

\begin{figure*}
\includegraphics[width=.95\textwidth]{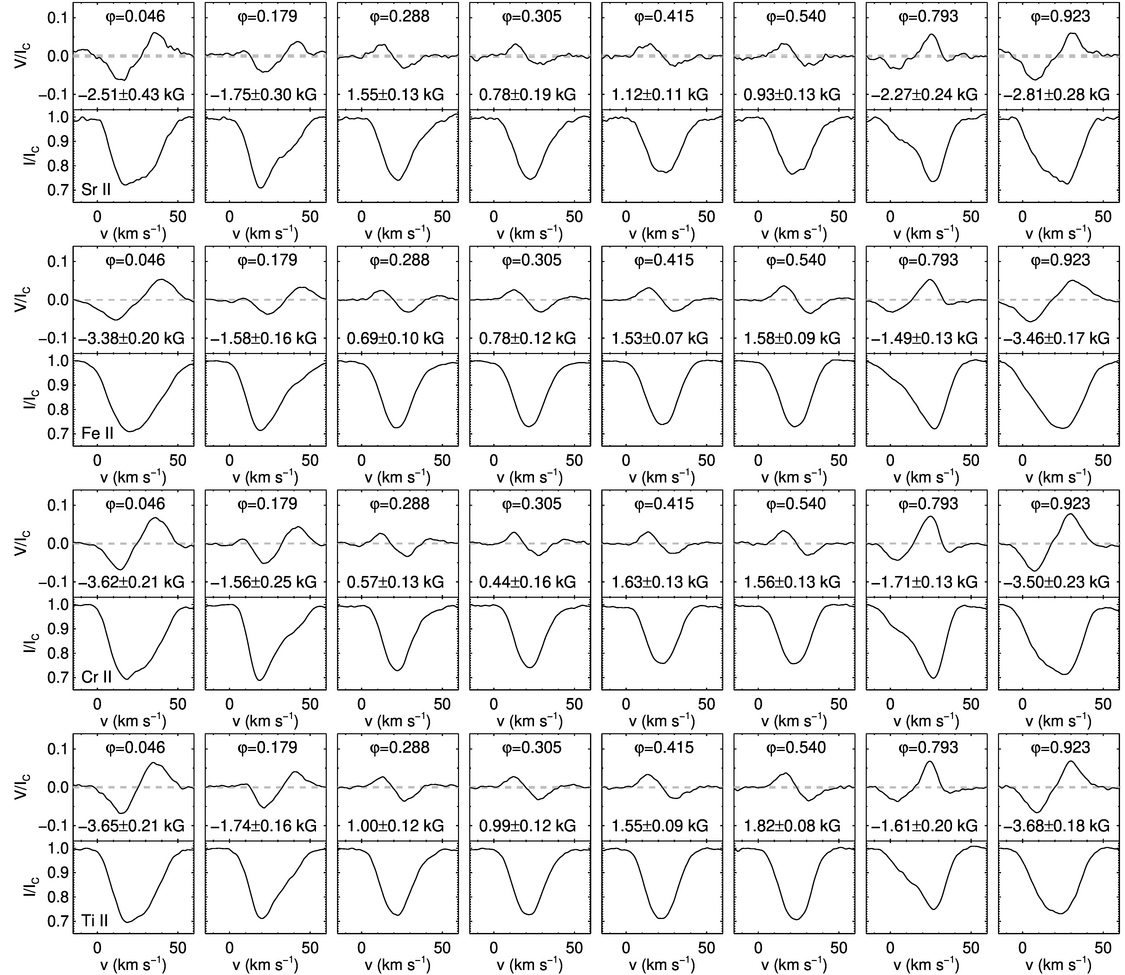}
\caption{Four panels present HARPS LSD Stokes~$I$ and Stokes~$V$ spectra of HD\,57372  employing line masks
with \ion{Sr}{ii}, \ion{Ti}{ii}, \ion{Cr}{ii}, \ion{Fe}{ii} lines, and ion{Sr}{ii} lines.
The dashed lines in the Stokes~$V$ panels indicate the mean uncertainties, which for the presented line masks are of 
the order of the thickness of these lines. 
\label{afig:IVNiron}}
\end{figure*}

\begin{figure*}
\includegraphics[width=\textwidth]{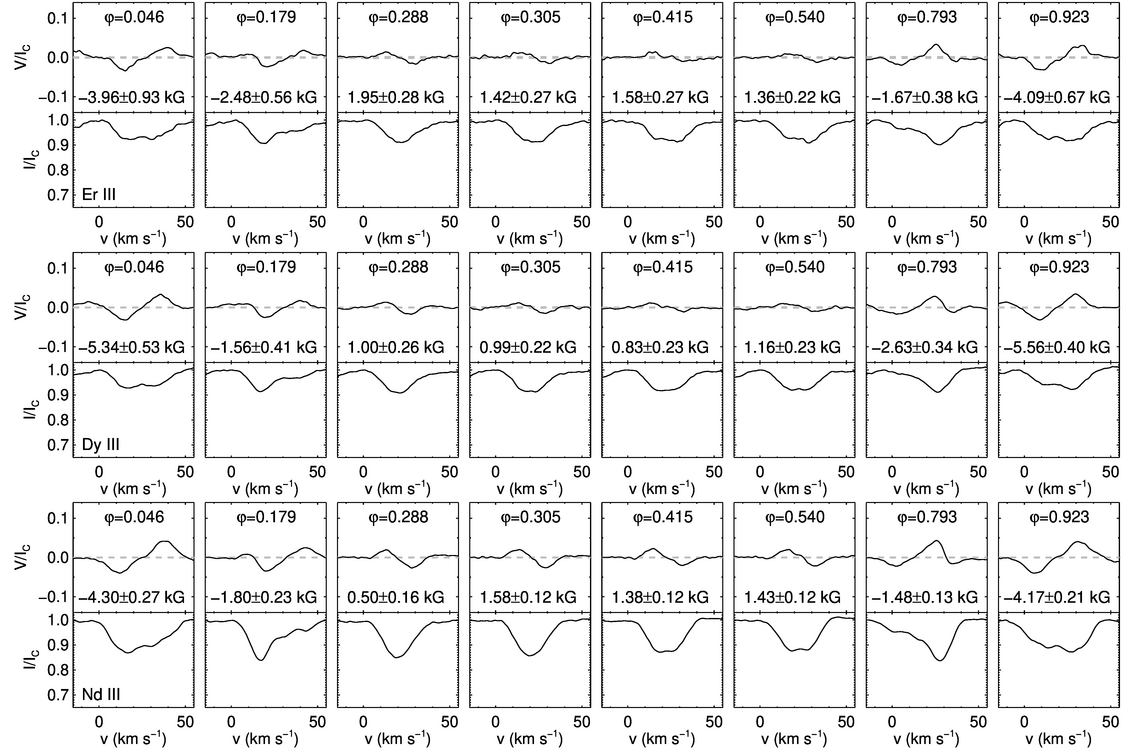}
\caption{As Fig.~\ref{afig:IVNiron}, but for REEs
  \ion{Nd}{iii}, \ion{Dy}{iii}, and \ion{Er}{iii}.
  \label{afig:IVNree}}
\end{figure*}

\begin{figure*}
\includegraphics[width=\textwidth]{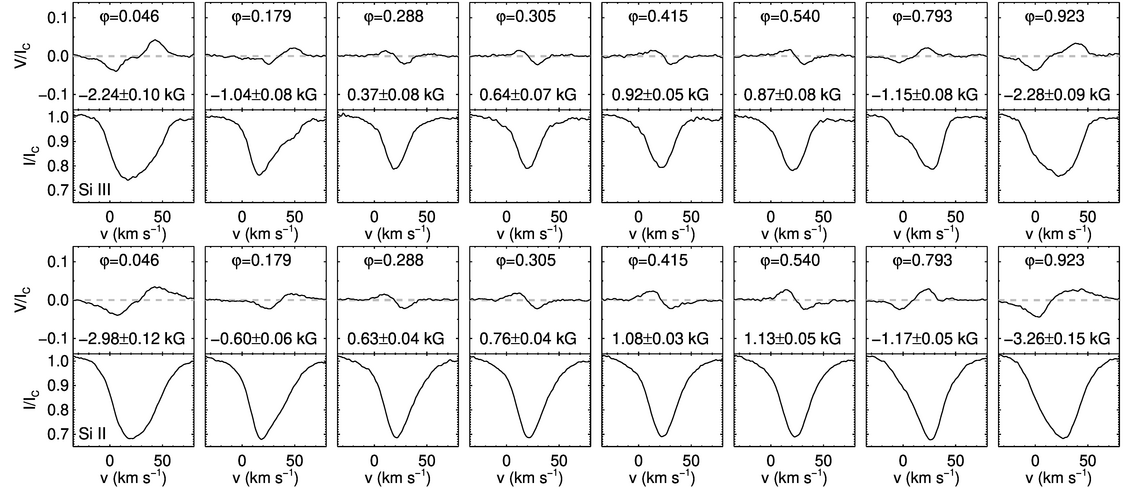}
\caption{As Fig.~\ref{afig:IVNiron}, but for \ion{Si}{ii} and \ion{Si}{iii}.
\label{afig:IVNSi}}
\end{figure*}

LSD Stokes~$I$, and $V$ profiles calculated for the separate masks with 
\ion{Ti}{ii}, \ion{Cr}{ii}, \ion{Fe}{ii}, and \ion{Sr}{ii}, are presented
in Fig.~\ref{afig:IVNiron}.
The results of the LSD technique applied for the lines masks constructed for rare earth
elements \ion{Nd}{iii}, \ion{Dy}{iii}, and
\ion{Er}{iii} are presented in Fig.~\ref{afig:IVNree}.
The comparison between Stokes~$I$ and $V$ profiles calculated for
once and twice ionised Si lines is shown in Fig.~\ref{afig:IVNSi}. As diagnostic null
spectra for all measurements appear flat and featureless, they are not presented
in these figures.
The dashed lines in Stokes~$V$ panels indicate the mean uncertainty,
which in numerous cases is not larger than the thickness of these lines.

\begin{table*}
\begin{center}
\caption{
LSD mean longitudinal magnetic field measurements for different line masks.
\label{tab:bz}
}
\begin{tabular}{lr r@{$\pm$}l r@{$\pm$}l r@{$\pm$}l r@{$\pm$}l r@{$\pm$}l
r@{$\pm$}l r@{$\pm$}l r@{$\pm$}l}
\hline
\multicolumn{1}{c}{Mask} &
\multicolumn{1}{c}{$\overline{g}_{\rm eff}$} &
\multicolumn{2}{c}{$\left< B_{\rm z}\right>$} &
\multicolumn{2}{c}{$\left< B_{\rm z}\right>$} &
\multicolumn{2}{c}{$\left< B_{\rm z}\right>$} &
\multicolumn{2}{c}{$\left< B_{\rm z}\right>$} &
\multicolumn{2}{c}{$\left< B_{\rm z}\right>$} &
\multicolumn{2}{c}{$\left< B_{\rm z}\right>$} &
\multicolumn{2}{c}{$\left< B_{\rm z}\right>$} &
\multicolumn{2}{c}{$\left< B_{\rm z}\right>$} \\
\multicolumn{1}{c}{} &
\multicolumn{1}{c}{} &
\multicolumn{2}{c}{(kG)} &
\multicolumn{2}{c}{(kG)} &
\multicolumn{2}{c}{(kG)} &
\multicolumn{2}{c}{(kG)} &
\multicolumn{2}{c}{(kG)} &
\multicolumn{2}{c}{(kG)} &
\multicolumn{2}{c}{(kG)} &
\multicolumn{2}{c}{(kG)} \\
\hline
\multicolumn{1}{c}{$\varphi_{\rm{rot}}$} &
\multicolumn{1}{c}{} &
\multicolumn{2}{c}{0.046} &
\multicolumn{2}{c}{0.179} &
\multicolumn{2}{c}{0.288} &
\multicolumn{2}{c}{0.305} &
\multicolumn{2}{c}{0.415} &
\multicolumn{2}{c}{0.540} &
\multicolumn{2}{c}{0.793} &
\multicolumn{2}{c}{0.923} \\
\hline
All (87)           & 1.14 & $-$3.32 & 0.09 & $-$1.43 & 0.08 & 0.64 & 0.06 & 0.79 & 0.06 & 1.43 & 0.04 & 1.43 & 0.04 & $-$1.73 & 0.06 & $-$3.22 & 0.10  \\
\ion{Ti}{ii} (10)  & 1.08 & $-$3.66 & 0.21 & $-$1.74 & 0.16 & 1.00 & 0.12 & 0.99 & 0.12 & 1.55 & 0.09 & 1.82 & 0.08 & $-$1.61 & 0.20 & $-$3.68 & 0.18  \\
\ion{Cr}{ii} (9)   & 1.05 & $-$3.62 & 0.21 & $-$1.56 & 0.25 & 0.58 & 0.13 & 0.44 & 0.16 & 1.63 & 0.13 & 1.56 & 0.13 & $-$1.71 & 0.13 & $-$3.50 & 0.23  \\
\ion{Fe}{ii} (22)  & 1.04 & $-$3.38 & 0.20 & $-$1.58 & 0.16 & 0.69 & 0.10 & 0.78 & 0.12 & 1.53 & 0.07 & 1.58 & 0.09 & $-$1.49 & 0.13 & $-$3.46 & 0.18  \\
\ion{Sr}{ii} (3)   & 1.20 & $-$2.51 & 0.43 & $-$1.75 & 0.30 & 1.55 & 0.13 & 0.78 & 0.19 & 1.12 & 0.11 & 0.93 & 0.13 & $-$2.27 & 0.24 & $-$2.81 & 0.28  \\
\ion{Nd}{iii} (14) & 1.18 & $-$4.30 & 0.27 & $-$1.80 & 0.23 & 0.50 & 0.16 & 1.59 & 0.12 & 1.38 & 0.12 & 1.43 & 0.12 & $-$1.48 & 0.13 & $-$4.17 & 0.21  \\
\ion{Dy}{iii} (11) & 1.15 & $-$5.34 & 0.53 & $-$1.56 & 0.41 & 1.00 & 0.26 & 0.99 & 0.22 & 0.83 & 0.23 & 1.16 & 0.23 & $-$2.63 & 0.34 & $-$5.56 & 0.40  \\
\ion{Er}{iii} (5)  & 1.18 & $-$3.96 & 0.93 & $-$2.48 & 0.56 & 1.95 & 0.28 & 1.42 & 0.27 & 1.58 & 0.27 & 1.36 & 0.22 & $-$1.67 & 0.38 & $-$4.09 & 0.67  \\
\ion{Si}{ii} (10)  & 1.19 & $-$2.98 & 0.12 & $-$0.60 & 0.06 & 0.63 & 0.04 & 0.76 & 0.04 & 1.08 & 0.03 & 1.13 & 0.05 & $-$1.17 & 0.05 & $-$3.26 & 0.15  \\
\ion{Si}{iii} (3)  & 1.83 & $-$2.24 & 0.10 & $-$1.04 & 0.08 & 0.37 & 0.08 & 0.64 & 0.07 & 0.92 & 0.05 & 0.87 & 0.08 & $-$1.15 & 0.08 & $-$2.28 & 0.09  \\
\hline
\end{tabular}
\end{center}
Notes:
In the first column we present the number of lines used in each mask followed by the average Land\'e factors
and the field measurements in kG.
All measurements are definite detections.
\end{table*}

\end{appendix}

\end{document}